\documentclass[sn-mathphys,Numbered]{sn-jnl}

\usepackage{graphicx}%
\usepackage{multirow}%
\usepackage{amsmath,amssymb,amsfonts}%
\usepackage{amsthm}%
\usepackage{mathrsfs}%
\usepackage[title]{appendix}%
\usepackage{xcolor}%
\usepackage{textcomp}%
\usepackage{manyfoot}%
\usepackage{booktabs}%
\usepackage{algorithm}%
\usepackage{algorithmicx}%
\usepackage{algpseudocode}%
\usepackage{listings}%
\usepackage[T1]{fontenc} 
\usepackage[utf8]{inputenc}
\usepackage{siunitx}
\usepackage{url}
\usepackage{tensor}
\usepackage{float}
\usepackage{mathtools}
\usepackage{amssymb}
\usepackage{amsmath}
\usepackage{makecell,tabularx,multirow}
\usepackage{color, colortbl}
\usepackage{adjustbox}
\usepackage[pages=some]{background}
\usepackage{hyperref}
\usepackage{accents}

\usepackage[font=footnotesize]{caption}

\hypersetup{
    colorlinks=true,
    linkcolor=blue,
    filecolor=magenta,
    citecolor=red
}
\usepackage{comment}
\usepackage[normalem]{ulem}

\usepackage[caption=false]{subfig}

\newcommand{\udt}[3]{#1^{#2}_{\phantom{#2}#3}}
\newcommand{\udut}[4]{#1^{#2\phantom{#3}#4}_{\phantom{#2}#3\phantom{#4}}}

\newcommand{\dut}[3]{#1_{#2}^{\phantom{#2}#3}}
\newcommand{\dudt}[4]{#1_{#2\phantom{#3}#4}^{\phantom{#2}#3}}

\newcommand{\lc}[1]{\accentset{\circ}{#1}}

\DeclareMathOperator{\sech}{sech}

\newcolumntype{C}{>{\centering\arraybackslash}X}
\newcolumntype{x}[1]{>{\centering\arraybackslash\hspace{0pt}}p{#1}}

\begin{document}

\title[\textbf{Cosmic growth in \texorpdfstring{$f(T)$}{} teleparallel gravity}]{\textbf{Cosmic growth in \texorpdfstring{$f(T)$}{} teleparallel gravity}}


\author[1,2,3]{\fnm{Salvatore} \sur{Capozziello}}\email{capozziello@na.infn.it}

\author*[4,5]{\fnm{Maria} \sur{Caruana}}\email{maria.caruana.16@um.edu.mt}

\author[4,5]{\fnm{Gabriel} \sur{Farrugia}}\email{gfarr02@um.edu.mt}

\author[4,5]{\fnm{Jackson} \sur{Levi Said}}\email{jackson.said@um.edu.mt}

\author[6]{\fnm{Joseph} \sur{Sultana}}\email{joseph.sultana@um.edu.mt}

\affil[1]{\orgdiv{Dipartimento di Fisica ``E. Pancin''}, \orgname{Universit\`a degli Studi di Napoli, ``Federico II'',  Complesso Universitario Monte S. Angelo}, \street{Via Cinthia 9 Edificio G}, \postcode{80126} \city{Napoli}, \country{Italy}}

\affil[2]{\orgdiv{Istituto Nazionale di Fisica Nucleare (INFN)}, \orgname{Sezione di Napoli Complesso Universitario Monte S. Angelo},\street{Via Cinthia 9 Edificio G}, \postcode{80126}\city{Napoli}, \country{Italy}}

\affil[3]{\orgdiv{Scuola Superiore Meridionale},\street{Largo San Marcellino 10}, \postcode{80138} \city{Napoli}, \country{Italy}}

\affil[4]{\orgdiv{Institute of Space Sciences and Astronomy}, \orgname{University of Malta}, \country{Malta}, \postcode{MSD 2080}}

\affil[5]{\orgdiv{Department of Physics}, \orgname{University of Malta}, \country{Malta}, \postcode{MSD 2080}}

\affil[6]{\orgdiv{Department of Mathematics}, \orgname{University of Malta}, \city{Msida}, \country{Malta}}


\abstract{Physical evolution of cosmological models can be tested by using expansion data, while  growth history of these models  is capable of testing  dynamics of the  inhomogeneous parts of  energy density. The growth factor, as well as its growth index, gives a clear indication of the performance of cosmological models in the regime of structure formation of  early Universe. In this work, we explore the growth index in several leading $f(T)$ cosmological models, based on a specific class of teleparallel gravity theories. These have become prominent in the literature and lead to other formulations of teleparallel gravity. Here we adopt a generalized approach by obtaining the M\'{e}sz\'{a}ros equation without immediately imposing the subhorizon limit, because  this assumption could lead to over-simplification. This approach gives avenue to study at which $k$ modes the subhorizon limit starts to apply. We obtain numerical results for the growth factor and growth index for a variety of data set combinations for each $f(T)$ model.}





\maketitle

\section{Introduction}\label{sec:intro}
The $\Lambda$CDM concordance model continues to agree with specific observational measurements at an extreme level of confidence at all cosmic scales \cite{Misner:1974qy,Clifton:2011jh}. In this paradigm, cold dark matter (CDM) acts as an agent to connect large scale structures on cosmic scales \cite{Peebles:2002gy,Baudis:2016qwx,Bertone:2004pz}, while dark energy is embodied in the cosmological constant \cite{Copeland:2006wr} as a driver for late cosmic acceleration. On the other hand, despite great efforts, the cosmological constant continues to have internal consistency issues \cite{Weinberg:1988cp} and falsifiable models of dark matter remain elusive in observations \cite{Gaitskell:2004gd}. In addition, there are other important anomalies in the $\Lambda$CDM setting of cosmic evolution such as the lithium problem \cite{DiBari:2013dna} in Big Bang nucleosynthesis measurements, as well as large scale cosmic microwave background (CMB) anomalies ranging from the lack of large-angle CMB temperature correlations \cite{doi:10.1073/pnas.90.11.4766} to asymmetries in full-sky measurements \cite{2020ApJS..247...69E,2007ApJ...660L..81E}, as well as other open questions \cite{Aghanim:2018eyx,Perivolaropoulos:2021jda}. These have been long-standing issues, but more recently, the effectiveness in describing combinations of data sets has been called into question through the so-called cosmic tensions issue \cite{DiValentino:2020vhf,DiValentino:2020zio,DiValentino:2020vvd,DiValentino:2020srs}. This is a statistical tension between direct measurements of cosmic expansion in the late Universe \cite{Riess:2019cxk,Anderson:2023aga,Wong:2019kwg}, and model-dependent measurements using data from the early Universe \cite{Aghanim:2018eyx,ACT:2023kun,Schoneberg:2022ggi}. It is possible that this is an artifact of the measurement techniques \cite{Riess:2020sih,Pesce:2020xfe,deJaeger:2020zpb,Capozziello:2023ewq}, but the range of  appearance of these tensions gives ground on which to consider more seriously new physical models of cosmology \cite{Abdalla:2022yfr,Bernal:2016gxb}.

The combination of challenges to the $\Lambda$CDM model has prompted a reevaluation of alternative formulations of several ideas in formulating different cosmological models \cite{Sotiriou:2008rp,Clifton:2011jh,CANTATA:2021ktz,Krishnan:2020vaf,Colgain:2022nlb,Malekjani:2023dky,Ren:2022aeo,Dainotti:2021pqg,Addazi:2021xuf,Schoneberg:2021qvd}. Taking the early- and late-Universe paradigm that comes from the cosmic tensions problem itself, there have been several models that modify standard cosmology in these separate sectors individually. Solutions in the late time regime of the Universe depend on changing the recent cosmic evolution without altering the early Universe. These range from decaying models of dark matter \cite{Anchordoqui:2020djl} or dark energy \cite{Alam:2003rw}, sometimes with interaction terms \cite{Gariazzo:2021qtg, Piedipalumbo:2023dzg}, as well as running vacua models \cite{Sola:2016jky} and sign switching models \cite{Akarsu:2023mfb}. The combination of supernovae of Type Ia (SNIa), cosmic chronometer (CC), and baryonic acoustic oscillation (BAO) data severely limits possible modifications to cosmic evolution away from the $\Lambda$CDM model, unless one considers deviations from the underlying principles of standard cosmology \cite{Colgain:2023bge}. On the other end of the cosmic spectrum, early Universe proposals are mainly fixed on modifications that decrease the sound horizon since this infers a higher Hubble constant for a fixed CMB angular scale. Examples of this include early Universe dark energy models \cite{Poulin:2023lkg}, as well as more exotic neutrino models \cite{DiValentino:2017oaw}, among others. However, the required decrease in the sound horizon renders these models incompatible with the growth of large scale structures \cite{Jedamzik:2020zmd}. This may point to a global rescaling of the Hubble evolution of the Universe, which may indeed take the form of a modified theory of gravity that acts differently from General Relativity (GR) in some important ways \cite{Escamilla-Rivera:2019hqt}.

Over the years, several proposals for different classes of modified gravity  have been used to tackle a varied collection of physical problems \cite{CANTATA:2021ktz,Capozziello:2002rd,Capozziello:2011et,Clifton:2011jh,Nojiri:2010wj,Nojiri:2017ncd}. It is worth noticing that GR relies on Riemann geometry whose dynamics is based on   curvature derived from metric. This happened mainly due to the fact that it was the only mature enough geometric tool during the time that GR was being proposed \cite{e24c2fd0-bd80-346c-b6f7-f15690a270c6}. Over  the following decades, other geometric approaches have been duly developed such as torsional models which are embodied in teleparallel gravity (TG) \cite{Bahamonde:2021gfp,Aldrovandi:2013wha,Cai:2015emx,Krssak:2018ywd}. TG has gained enormous popularity in recent years and relies on the exchange of the curvature-based Levi-Civita connection $\lc{\Gamma}^{\sigma}{}_{\mu\nu}$ (over-circles denotes quantities based on  the Levi-Civita connection in this work) with the torsion-based teleparallel connection $\Gamma^{\sigma}{}_{\mu\nu}$. This picture may provide a more intuitive approach in considering new physics at the level of gravitational theory.

The teleparallel connection is curvature-free and satisfies metricity, as well as being the basis of a novel framework for gravitation \cite{Aldrovandi:2013wha}. On the other hand, a particular combination of gravitational contractions exists  producing the teleparallel equivalent of General Relativity (TEGR), which is dynamically equivalent to GR at the level of the classical field equations (see \cite{Capozziello:2022zzh} for a discussion). This means that both GR and TEGR  have identical predictions at the level of astrophysical and cosmological phenomenology, but differences may arise when IR completions are considered \cite{Mylova:2022ljr}. Here, the action is  linear in the torsion scalar $T$. Now, modifications to TG through the TEGR action can be explored using the same rationale as in regular modified gravity in the regime of GR. One popular approach to generalizing TEGR is  $f(T)$ teleparallel gravity \cite{Ferraro:2006jd,Ferraro:2008ey,Bengochea:2008gz,Linder:2010py,Chen:2010va,Basilakos:2013rua,Cai:2015emx,Bahamonde:2019zea,Paliathanasis:2017htk,Farrugia:2020fcu,Bahamonde:2021srr,Bahamonde:2020bbc} which is based on an arbitrary functional of the torsion scalar and produces generically second order equations of motion for all spacetimes. This has also been further extended in other directions such as with a scalar field \cite{Bahamonde:2019ipm,Capozziello:2023foy,Dialektopoulos:2021ryi,Bernardo:2021bsg,Bernardo:2021izq,Bahamonde:2021dqn,Bahamonde:2020cfv}, as well with other scalar modifications \cite{Capozziello:2016eaz,Bajardi:2023gkd}, as in regular curvature-based modified gravity.

$f(T)$ gravity has an intriguing perturbative structure \cite{Bahamonde:2021gfp,Cai:2015emx,Krssak:2018ywd}, while being relatively straightforward (being second order), it poses possible problems of strong coupling \cite{BeltranJimenez:2020fvy}. 

In this work, we are interested in exploring how the growth of large scale structures is affected by the arbitrary inclusion of torsion scalar in the gravitational action. While TEGR is dynamically equivalent to GR, modifications to selective $f(T)$ gravity models may play a role in different regimes of cosmology \cite{Capozziello:2015rda,Capozziello:2017uam,Capozziello:2017bxm,Benetti:2020hxp}. Together, these arguments may collectively address the problem of cosmic tensions and may enrich physics at other scales. We do this by first introducing TG and its background equations for a homogeneous and isotropic flat cosmology in Sec.~\ref{sec:f_T_cosmo_intro}. To probe the effects on the growth of large scale structures, we consider matter perturbations in Sec.~\ref{sec:pertubation}. Gravitational scalar perturbations couple to matter perturbations and so can be used to explore how modifications to the gravitational sector modify the growth evolution of matter in the formation of large scale structures. These equations of motion can be reformulated to give a matter perturbation equation, the so-called M\'{e}sz\'{a}ros equation, through which the evolution of matter perturbations can be suitably investigated at different horizon scales. We do this in Sec.~\ref{sec:meszaros_equation} where the growth factor is defined. The growth factor was first studied for $f(T)$ gravity in Ref.~\cite{Zheng:2010am} where the growth of structures can be slowed down due to weakening in gravity for some $f(T)$ models. This may address the high value of the $S_8$ parameter \cite{Planck:2018vyg,Heymans:2020gsg,DES:2021wwk,Dalal:2023olq}. The growth factor in TG has also been calculated for $f(T,\Theta)$ gravity in Ref.~\cite{Farrugia:2016pjh} where other matter properties were also incorporated into the calculation. The behaviour of growth factor and constant growth index value are obtained for different $f(T)$ models in Sec.~\ref{sec:models}. In Sec.~\ref{sec:conclu}, we compare our results with these works and give an outlook of our main results.

\section{\texorpdfstring{$f(T)$}{} gravity and cosmology} \label{sec:f_T_cosmo_intro}

The curvature-based theory of GR is constructed through the fundamental objects of the metric and the torsionless Levi-Civita connection $\lc{\Gamma}^{\lambda}_{\mu\nu}$~\cite{misner1973gravitation}. The GR geometric framework can be recast into the (metric) torsionful theory of teleparallel gravity based on the teleparallel connection $\Gamma^{\lambda}_{\nu\mu}$ constructed by the  tetrads $\udt{e}{A}{\mu}$ as dynamical fundamental objects  and the spin connection $\udt{\omega}{A}{B\mu}$~\cite{Aldrovandi:2013wha}. The tetrad acts as a link between the Minkowski (Latin indices) and general manifold (Greek indices)~\cite{Hehl:1976kj}
\begin{align}\label{eq:metric_def}
    g_{\mu\nu} = \eta_{AB}\,\udt{e}{A}{\mu}\,\udt{e}{B}{\nu}\,,
    \qquad \qquad\qquad
    \eta_{AB} = g_{\mu\nu}\,\dut{E}{A}{\mu}\,\dut{E}{B}{\nu}\,,
\end{align}
where the inverse tetrad $\dut{E}{A}{\mu}$ satisfies orthogonality conditions
\begin{align} \label{eq:tet_orthogonality}
    \udt{e}{A}{\mu}\,\dut{E}{B}{\mu} = \delta^{A}_{B}\,,
    \qquad \qquad\qquad\qquad
    \udt{e}{A}{\mu}\, \dut{E}{A}{\nu} = \delta^{\nu}_{\mu}\,.
\end{align}
Hence, the teleparallel connection is defined by the TG variables as
\begin{align}~\label{eq:tele_connection}
    \Gamma^{\lambda}_{\nu\mu} = \dut{E}{A}{\lambda} \left( \partial_{\mu} \udt{e}{A}{\nu} + \udt{\omega}{A}{B\mu} \udt{e}{B}{\nu}\right)\,.
\end{align}
It leads to a vanishing curvature and satisfies non-metricity~\cite{Cai:2015emx,Krssak:2018ywd,Capozziello:2022zzh}. The TG inertial spin connection is flat such that it satisfies~\cite{Bahamonde:2021gfp} 
\begin{align}\label{eq:flat_spin_connection}
    \partial_{[\mu}\udt{\omega}{A}{|B|\nu]} + \udt{\omega}{A}{C[\mu} \, \udt{\omega}{C}{|B|\nu]} = 0\,,
\end{align}
where square brackets denote the antisymmetric operator, and embodies Local Lorentz Transformation (LLT) invariance such that
\begin{align}\label{eq:spin_connection}
    \udt{\omega}{A}{B\mu} = \udt{\Lambda}{A}{C}\,\partial_{\mu}\dut{\Lambda}{B}{C}\,.
\end{align}
Here $\udt{\Lambda}{A}{B}$ denotes Lorentz boosts and rotations~\cite{Aldrovandi:2013wha}. For any metric, an infinite number of tetrad choices could emerge. Hence, the spin connection becomes an integral part to ensure inertial effects are accounted for a covariant theory~\cite{Krssak:2015oua}. Nevertheless, there exists a Lorentz frame for which the spin connection vanishes leading to the application of the Weitzenb\"{o}ck gauge: $\udt{\omega}{A}{B\mu} = 0$~\cite{Weitzenbock:1923efa,Krssak:2018ywd}.

While the Riemann tensor, constructed on the Levi-Civita connection, describes the curvature (and then dynamical) properties of GR, its TG analog becomes the torsion tensor~\cite{Aldrovandi:2004db,Krssak:2018ywd}
\begin{align} \label{eq:torsion_tensor}
    \udt{T}{A}{\mu\nu} = \Gamma^{A}_{\nu\mu} - \Gamma^{A}_{\mu\nu}\,,
\end{align}
which convariantly transforms under LLTs and diffeomorphisms~\cite{Bahamonde:2021gfp}. Thus, the difference between the Levi-Civita and Weitzenb\"{o}ck connections is given by the contorsion tensor~\cite{Aldrovandi:2004db}
\begin{align}\label{eq:contorsion_tensor}
    \udt{K}{\lambda}{\mu\nu} = \Gamma^{\lambda}_{\mu\nu} - \lc{\Gamma}^{\lambda}_{\mu\nu} = -\frac{1}{2}\left(\udt{T}{\lambda}{\mu\nu} - \dudt{T}{\mu}{\lambda}{\nu} - \dudt{T}{\nu}{\lambda}{\mu}\right)\,,
\end{align}
and the potential relation to the gravitational energy-momentum is represented by the superpotential tensor~\cite{Nesseris:2013jea,Aldrovandi:2004db,Koivisto:2019ggr}
\begin{align}\label{eq:superpotential_tensor}
    \dut{S}{\lambda}{\mu\nu} = \frac{1}{2} \left( 
\udt{K}{\mu\nu}{\lambda} + \delta^{\mu}_{\lambda} \udt{T}{\alpha\nu}{\alpha} - \delta^{\nu}_{\lambda} \udt{T}{\alpha\mu}{\alpha} \right)\,.
\end{align}
Thus, the contraction between the torsion tensor~\eqref{eq:torsion_tensor} and superpotential tensor~\eqref{eq:superpotential_tensor} gives rise to the torsion scalar~\cite{Cai:2015emx}
\begin{align}\label{eq:torsion_scalar}
    T = \udt{T}{\lambda}{\mu\nu} \dut{S}{\lambda}{\mu\nu} = \frac{1}{4} T^{\lambda\mu\nu}T_{\lambda\mu\nu} + \frac{1}{2}T^{\lambda\mu\nu}T_{\mu\lambda\nu} - \dut{T}{\lambda\mu}{\lambda} \udt{T}{\nu\mu}{\nu}\,.
\end{align}
Hence, a dynamical equivalence relation can be derived between the Ricci scalar $\lc{R}$, constructed through the Levi-Civita connection, and the torsion scalar $T$ built by the Weitzenb\"{o}ck connection~\cite{Bahamonde:2015zma,Hehl:1976kj}
\begin{align} \label{eq:R_T_relation}
    R = \lc{R}+T-B = 0\,,
\end{align}
since $R$ vanishes in the metric-teleparallel formalism. Here
\begin{align}\label{eq:boundary_term}
    B = \tfrac{2}{e} \partial_{\mu}\left(e \udut{T}{\lambda}{\lambda}{\mu}\right) = 2 \lc{\nabla}_{\mu} \udut{T}{\lambda}{\lambda}{\mu}\,,
\end{align}
is the boundary term which encapsulates the fourth order derivative contributions of the field equations~\cite{Bahamonde:2015zma,Capozziello:2019msc}, typically accounted for within the Ricci scalar $\lc{R}$~\cite{Sotiriou:2008rp,Capozziello:2011et} and $e = \text{det}(\udt{e}{A}{\mu}) = \sqrt{-g}$. The boundary term $B$ leads to the teleparallel equivalence to General Relativity (TEGR) with the action constructed by a linear contribution of $T$~\cite{Hehl:1994ue,BeltranJimenez:2019esp} as seen in Eq.~\eqref{eq:R_T_relation} such that 
\begin{align}\label{eq:TEGR_action}
    \mathcal{S} = -\frac{1}{2\kappa^{2}} \int d^{4}x\, e\, T + \int d^{4}x\, e\, \mathcal{L}_m\,,
\end{align}
where $\kappa^{2} = 8 \pi G$ and $\mathcal{L}_{m}$ is the matter Lagrangian.
The extension of the Einstein-Hilbert action to the more general form of $f(\lc{R})$~\cite{Capozziello:2011et,Sotiriou:2008rp}, can be replicated by broadening on the TEGR action to obtain $f(T,B)$ gravity \cite{Bahamonde:2016grb,Capozziello:2019msc}. 
In this work, we consider the  $f(T)$ extension such that the action becomes~\cite{Ferraro:2008ey,Ferraro:2006jd,Bengochea:2008gz,Chen:2010va} 
\begin{align}\label{eq:f_T_action} 
    \mathcal{S} = -\frac{1}{2\kappa^{2}}\int d^{4}x\, e\,\left( T + f(T)\right) + \int d^{4}x\, e\, \mathcal{L}_m\,.
\end{align}
Through the weakening of Lovelock's theorem in TG, the equations of motion in $f(T)$ remain second order~\cite{Lovelock:1971yv,Gonzalez:2015sha,Bahamonde:2019shr}, unlike the fourth order equations of $f(\lc{R}) = f(-T+B)$, thus resulting in a naturally Gauss-Ostrogadsky ghost free form~\cite{Krssak:2018ywd,ortín2004}.

Let us consider a spatially flat Friedman-Lema\^{i}tre-Robertson-Walker (FLRW) metric in terms of Cartesian coordinates~\cite{Krssak:2018ywd} 
\begin{align}\label{eq:FLRW_metric}
    \mathrm{d}s^{2} = \mathrm{d}t^{2} - a(t)^{2}(\mathrm{d}x^{2} + \mathrm{d}y^{2} + \mathrm{d}z^{2})\,,
\end{align}
where $a(t)$ is the scale factor. The Weitzenb\"{o}ck gauge compatible tetrad is chosen to be
\begin{align}\label{eq:tetrad_choice}
    \udt{e}{A}{\mu} = \text{diag}\left(1,a(t),a(t),a(t)\right)\,,
\end{align}
such that the tetrad determinant is $e = a^{3}$. The torsion scalar can be written in terms of the Hubble parameter $H(t) = \tfrac{\dot{a}}{a}$ (overhead dot represents derivative with respect to time) as
\begin{align} \label{eq:Torsion_inH}
    T = -6H^{2}\,.
\end{align}
Taking the variation with respect to the tetrad yields field equations~\cite{Zheng:2010am}  
\begin{align}\label{eq:variation_tetrad}
    \dut{W}{C}{\alpha}: \qquad 
 \frac{1}{4}\dut{E}{C}{\alpha} \left(T + f \right) &- \partial_{\rho}T\, f_{TT}\, \dut{E}{C}{\lambda} \dut{S}{\lambda}{\rho\alpha} \nonumber \\ \qquad & - \left(1+f_{T}\right) \left(\dut{E}{C}{\mu} \udt{T}{\lambda}{\mu\nu} \dut{S}{\lambda}{\alpha\nu} + e^{-1} \partial_{\rho}\left(e\, \dut{E}{C}{\lambda} \dut{S}{\lambda}{\rho\alpha} \right)\right)  = \frac{\kappa^{2}}{2}\dut{\theta}{C}{\alpha} \,,
\end{align}
where $T$ subscripts denotes partial derivative with respect to the torsion scalar and the energy-momentum tensor $\dut{\theta}{C}{\gamma}$ for a perfect fluid is given by~\cite{Zheng:2010am,Farrugia:2016pjh}
\begin{align} \label{eq:energy_momentum_tensor}
    \dut{\theta}{C}{\gamma} = e^{-1} \frac{\partial(e\,\mathcal{L}_{m})}{\partial \udt{e}{C}{\gamma}} 
    = -\begin{bmatrix}
    \rho(t) & 0 \\
    0 & -\delta^{i}_{j} p(t)\,
    \end{bmatrix},
\end{align}
where $\rho$ is the matter energy density and $p$ is the matter pressure, and where the negative sign follows from the sign of the Lagrangian. By considering the components of Eq.~\eqref{eq:variation_tetrad}, the background cosmological equations of motion result in~\cite{Linder:2010py}
\begin{align}
   \label{eq:background_friedmann_1} \dut{W}{0}{0}: \,\,\,\,\qquad 2 \kappa^{2} \rho &= 6H^{2} + f + 12 H^{2} f_{T}\,, \\
    \label{eq:background_friedmann_2} \dut{W}{i}{i}: \qquad -2\kappa^{2}p &= f - 6H^{2} + 4(3H^{2} + \dot{H})(1+f_{T}) - 48 H^{2} \dot{H}f_{TT}\,.
\end{align}
By drawing a comparison with the analog GR field equations, the modified contributions can be identified to be 
\begin{align}
    \label{eq:modified_friedmann_1}
    2 \kappa^{2} \rho_{\text{DE}} &:= - f - 12 H^{2} f_{T}\,,\\
    \label{eq:modified_friedmann_2}
    -2\kappa^{2}  p_{\text{DE}} &:= 2\kappa^{2}\rho_{\text{DE}} - 4\dot{H} f_{T} + 48 H^{2} \dot{H} f_{TT}\,,
\end{align}
where $\rho_{\text{DE}}$ and $p_{\text{DE}}$ are the effective dark energy density and pressure, respectively. Thus, the effective equation of state is given by
\begin{align}\label{eq:EoS}
    \omega_{\text{DE}} &= \frac{p_{\text{DE}}}{\rho_{\text{DE}}} = -1 -4 \frac{\dot{H}f_{T} - 12H^{2}\dot{H}f_{TT}}{f + 12H^{2}f_{T}}\,,
\end{align}
provided $f \neq - 12H^{2}f_{T}$~\cite{Zheng:2010am}.

The conservation of energy-momentum tensor is given by~\cite{misner1973gravitation}
\begin{align}\label{eq:conservation_energy_momentum}
    \lc{\nabla}_{\mu}\udt{\theta}{\mu}{\nu} =\partial_{\mu}\udt{\theta}{\mu}{\nu} + \lc{\Gamma}^{\mu}_{\mu\alpha} \udt{\theta}{\alpha}{\nu} - \lc{\Gamma}^{\alpha}_{\mu\nu} \udt{\theta}{\mu}{\alpha} = 0\,,
\end{align}
where $\lc{\nabla}$ is the covariant derivative with respect to the Levi-Civita connection. Once again, by considering the components, the background continuity equation is
\begin{align} \label{eq:background_continuity}
    \dot{\rho}+3H(\rho+p) = 0\,.
\end{align}
The $\Lambda$CDM limit can be obtained by setting $f(T)$ to a constant so that the equation of state turns out to be $\omega_{\text{DE}} = -1$. By Eq.~\eqref{eq:modified_friedmann_1},  it is clear that the effective energy density becomes equivalent to a constant, also referred to as the cosmological constant $\Lambda$.

\section{First order scalar perturbations}\label{sec:pertubation}

The first order scalar perturbations, which couple to the matter density, represent the inhomogeneities in the evolution of the Universe. The first order scalar perturbation of the FLRW metric in the longitudinal (Newtonian) gauge is given  by~\cite{Zheng:2010am,Farrugia:2016pjh}
\begin{align}\label{eq:FLRW_metric_longitudinal}
    \mathrm{d}s^{2} = (1+2\varphi) \mathrm{d}t^{2} - a(t)^{2}(1-2\psi)\delta_{ij}\mathrm{d}x^{i}\mathrm{d}x^{j}\,,
\end{align}
where $\varphi$ and $\psi$ are scalar modes, and the off-diagonal terms vanish. It should be noted, that the decoupling of the scalar, vector and tensor modes makes it possible to treat the sectors separately. The general perturbed tetrad choice about the background $\udt{\bar{e}}{A}{\mu}$ in Eq.~\eqref{eq:tetrad_choice} is given by~\cite{Zheng:2010am,Farrugia:2016pjh}
\begin{align}\label{eq:tetrad_pert_expression}
    \udt{e}{A}{\mu} = (\delta^{A}_{B} + \dut{X}{B}{A}) \udt{\bar{e}}{B}{\mu}\,, \qquad\qquad \left|\dut{X}{B}{A} \right| \ll 1\,,
\end{align}
where
\begin{align} \label{eq:tetrad_firstorder}
    \dut{X}{B}{A} = \begin{bmatrix}
        \varphi &&& -\delta^{I}_{i}\partial^{i} \omega\\
        \delta^{i}_{I}\partial_{i}\bar{w} &&& -\delta^{I}_{i} \delta^{j}_{J} (\delta^{i}_{j} \psi + \partial^{i}\partial_{j} h +\udt{\epsilon}{i}{jk}\partial^{k}\sigma)
    \end{bmatrix}\,,
\end{align}
wherein $\omega$, $\bar{\omega}$ and $h$ are additional scalar modes, and $\sigma$ is the pseudo-scalar and $\epsilon_{ijk}$ is the spatial Levi-Civita symbol. Thus, substituting into Eq.~\eqref{eq:tetrad_pert_expression} and Eq.~\eqref{eq:metric_def}, one can obtain the perturbed tetrad and metric tensor, respectively, 
\begin{align}\label{eq:pert_tetrad_general}
    \udt{e}{A}{\mu} &= \begin{bmatrix}
        1+\varphi &&& a \partial_{i}\bar{\omega}\\
        - \delta^{I}_{i} \partial^{i}\omega &&& a 
        \delta^{Ii} \left[\delta_{ij}(1-\psi) - \partial_{i}\partial_{j}h + \epsilon_{ijk} \partial^{k}\sigma \right]
    \end{bmatrix}\,,\\ \nonumber \\
    \label{eq:pert_metric_general}
    g_{\mu\nu} &= \begin{bmatrix}
        1+2\varphi & a\partial_{i}(\omega + \bar{\omega})\\
        a\partial_{i}(\omega + \bar{\omega}) & -a^{2}\left[(1-2\psi)\delta_{ij} - 2\partial_{i}\partial_{j}h\right]
    \end{bmatrix}\,.
\end{align}
Drawing a comparison between the general metric in Eq.~\eqref{eq:pert_metric_general} and the longitudinal gauge version in Eq.~\eqref{eq:FLRW_metric_longitudinal} implies that the longitudinal gauge conditions are given by $\bar{\omega} = -\omega$ and $h = 0$. The additional modes of $\omega$ and $\sigma$ in the tetrad, albeit the former vanishes, due to the longitudinal gauge, and the latter naturally cancels out in the metric tensor, should still be implemented in the tetrad choice as they will play a role in the system of field equations~\cite{Zheng:2010am}. Additionally, since we are working within the Weitzenb\"{o}ck gauge, its application can be extended to first order such that contributions of the spin connection at the perturbative level can be set to vanish~\cite{Hohmann:2020vcv,Golovnev:2018wbh}. The energy-momentum tensor can be expressed in terms of the fluid four-velocity $u^{\mu}$~\cite{misner1973gravitation}
\begin{align}\label{eq:four_vel_energymomentum}
    \theta^{\mu\nu} = (\rho+p)u^{\mu}u^{\nu} - pg^{\mu\nu}  - \Pi^{\mu\nu}\,.
\end{align}
where we have introduced the anisotropic stress tensor $\Pi^{\mu\nu}$. Hence, the energy-momentum tensor can be expanded up to first order~\cite{Farrugia:2016pjh}
\begin{align}\label{eq:pert_energymomentum}
    \dut{\theta}{C}{\gamma} = -\begin{bmatrix}
        \rho+\delta\rho & (\rho+p)\partial^{i}v\\
        -a^2 \delta_{I}^{i}(\rho+p)\partial_{i}v & -\delta^{Ij} \left(\delta_{ij}(p+\delta p) + \partial_{i}\partial_{j}\pi^{\text{s}}\right)
    \end{bmatrix}\,,
\end{align}
where $v$ is the scalar component of the velocity vector $v^{i} = \frac{u^{i}}{u^{0}}$ and $\pi^{\text{s}}$ is the scalar component of the anisotropic stress wherein the properties $\udt{\Pi}{0}{0} = \udt{\Pi}{0}{i} = u^{\mu} \Pi_{\mu\nu} = 0$ are satisfied. It should be noted that all perturbation modes are taken to be temporal and spatial.

First order perturbation of Eq.~\eqref{eq:variation_tetrad} give rise to a system of fields equations when considering its components along with the background equations~\eqref{eq:background_friedmann_1} and ~\eqref{eq:background_friedmann_2} as follows
\begin{align}
    \label{eq:W00}
    \dut{W}{0}{0}: &\,&  0 &= \kappa^{2} \delta\rho + 6 H ( 1 + f_{T} - 12 H^{2}f_{TT} ) ( H\varphi + \dot{\psi}) + 2 a^{-2}(1 + f_{T})\partial^{2}\psi \nonumber \\ 
    &&& \qquad 
    - 24 a^{-1} H^{3}  f_{TT}\partial^{2}\omega\,,  \\
    \label{eq:W0i}
    \dut{W}{0}{i}: &\,&  0 &= \kappa^{2} a^{2} (\rho + p) \partial^{i}v + 2 (1+f_{T}) \partial^{i}(H \varphi + \dot{\psi}) - 24 H \dot{H} f_{TT} \partial^{i}\psi \,, \\
    \label{eq:Wi0}
    \dut{W}{i}{0}: &\,&  0 &= \kappa^{2} a^{2} (\rho + p) \partial_{i}v + 2 (1 + f_{T} - 12 H^{2} f_{TT}) \partial_{i}(H\varphi + \dot{\psi}) - 8 a^{-1} H^{2} f_{TT} \partial_{i} \partial^{2} \omega
    \,, \\
    \label{eq:Wij}
    \dut{W}{i}{j}: &\,&  0 &= (1+f_{T}) \partial_{i}\partial^{j} (\varphi - \psi) + 12 a H \dot{H} f_{TT} \partial_{i}\partial^{j}\omega + \kappa^{2} a^{2} \partial_{i}\partial^{j} \pi^{\text{s}}
    \,, \\
    \label{eq:Wii}
    \dut{W}{i}{i}: &\,&  0 &= + \kappa^{2} (3 \delta p - \partial^{2}\pi^{\text{s}}) + 2 a^{-2} (1+f_{T}) \partial^{2}(\varphi - \psi) - 6 H (1 + f_{T} - 12 H^{2} f_{TT}) (\dot{\varphi} + 3 \dot{\psi}) \nonumber \\
     &&& \quad
    -6 [
    2 \dot{H} (1+f_{T}) + 3 H^{2} (1 + f_{T} - 20 \dot{H} f_{TT}) - 36 H^{4} (f_{TT} - 4 \dot{H} f_{TTT}) ]  \varphi \nonumber \\
    &&& \quad 
    - 24 a^{-1}H [ -3 \dot{H} f_{TT} \partial^{2}\omega - 2 H^{2} (f_{TT} - 6 \dot{H} f_{TTT}) \partial^{2}\omega - H f_{TT} \partial^{2}\dot{\omega}]
    \nonumber \\ &&& \quad
    + 216 H ( \dot{H} f_{TT} - 4 H^{2} \dot{H} f_{TTT}) \dot{\psi} - 6 (1 + f_{T} - 12 H^{2}f_{TT}) \ddot{\psi}
     \,,
\end{align}
where $\partial^{2} = -\partial_{i}\partial^{i}$. Similarly, the conservation of energy-momentum tensor in Eq.~\eqref{eq:conservation_energy_momentum} is perturbed to obtain the first order continuity and velocity (Euler) equations by applying their background equation counterpart Eq.~\eqref{eq:background_continuity} as follows:
\begin{align}
    \label{eq:pert_continuity}
    0 &= \dot{\delta\rho} + (\rho + p) \left(\partial^{2}v - 3 \dot{\psi}\right) + 3H \left(\delta\rho + \delta p - \frac{\partial^{2}\pi^{\text{s}}}{3}\right)\,, \\
    \label{eq:pert_velocity}
    0 &= \partial_{i}\delta p - \partial_{i}\partial^{2}\pi^{\text{s}} + \left(\rho + p\right) \left( \partial_{i}\varphi + 2 a^{2} H \partial_{i}v + a^{2} \partial_{i}\dot{v}\right) + a^{2} \dot{p} \partial_{i}v\,.
\end{align}
Here, the evolution of the scalar perturbations can be represented by the gravitational perturbations equations~(\ref{eq:W00}-\ref{eq:Wii}) and conservation of energy-momentum tensor equations~(\ref{eq:pert_continuity}-\ref{eq:pert_velocity}).

\section{The M\'{e}sz\'{a}ros equation in the subhorizon limit} \label{sec:meszaros_equation}

To study the growth for a variety of models in $f(T)$ gravity, we are interested in the growth of the cold dark matter content, 
\begin{align}\label{eq:matter_condition}
    p = \delta p = 0\,,   
\end{align}
where we perform the analysis starting within the matter dominated epochs up to the current time, as radiation eras are suppressed during these epochs. The contribution of the anisotropic stress presented in Eq.~\eqref{eq:energy_momentum_tensor} represents the quadrupole contributions that play a role mainly in the radiation era and are small when considering cold dark matter content during the matter dominated era~\cite{Dodelson:2003ft}. Hence, $\pi^{\text{s}}$ is set to vanish for this analysis. In Ref.~\cite{Zheng:2010am}, the growth factor analysis was carried out through the immediate application of the subhorizon condition $k \gg a H$, where $k$ is the spatial Fourier transformation wave mode such that $\partial^{2} = k^{2}$. In this work, a more general approach is considered to investigate the $k$-dependency within the growth factor and growth index. As seen in Ref.~\cite{delaCruz-Dombriz:2008ium}, in the case of $f(\lc{R})$ gravity, it was shown that obtaining the M\'{e}sz\'{a}ros equation, through a system of equations within the subhorizon limit, could lead to some discrepancies when compared to the quasi-static result obtained by ignoring time derivatives of the potential~\cite{Tsujikawa:2007gd,DeFelice:2010aj-f(R)_theories}. This shows that, in general, the use of the subhorizon limit, despite simplifying the procedure, should be limited. Hence, the general procedure implemented in Ref.~\cite{delaCruz-Dombriz:2008ium} is followed to obtain a more general result than Ref.~\cite{Zheng:2010am}.

First and foremost, Fourier transformation on the spatial portions of Eqs~(\ref{eq:W00}-\ref{eq:pert_velocity}) is carried out. Hence, unless otherwise stated, variables presented from this point on are in their Fourier transform form. Next, we define the gauge invariant comoving fractional matter perturbation as
\begin{align}\label{eq:fractional_matter}
    \delta_{m} \equiv \frac{\delta\rho}{\rho }-3 a^{2} H v\,.
\end{align}
Furthermore, for the sake of simplicity, the velocity scalar mode $v$ is redefined as
\begin{align}\label{eq:velocity_redefinition}
    V = a^{2} v\,.
\end{align}
By considering these substitutions within the continuity equation~\eqref{eq:pert_continuity} and velocity equation~\eqref{eq:pert_velocity}, the two equations reduce to
\begin{align}
    \label{eq:continuity_2}
    0 &= \dot{\delta}_{m} - 3 \dot{\psi} + V \left(-\frac{k^{2}}{a^{2}} + 3 \dot{H} \right) + 3 H \dot{V}\,, \\
    \label{eq:velocity_2}
    0 &= \varphi + \dot{V}\,,
\end{align}
where $\pi^{\text{s}} = 0$. By solving Eq.~\eqref{eq:velocity_2} for $\dot{V}$ and substituting into Eq.~\eqref{eq:continuity_2} such that
\begin{align}\label{eq:continuity_final}
    V \left(\frac{k^{2}}{a^{2}} - 3 \dot{H}\right) = \dot{\delta}_{m} - 3 \left(H \varphi + \dot{\psi} \right)\,,
\end{align}
provides an expression for $V$. Additionally, Eq.~\eqref{eq:Wij} [$\dut{W}{i}{j}$] provides an expression for the $\omega$ scalar mode
\begin{align} \label{eq:omega_expression}
    \omega &= - \frac{\left(\varphi - \psi\right)\left(1 + f_{T}\right)}{12 a H \dot{H} f_{TT}}\,.
\end{align}
Hence, substituting Eqs~\eqref{eq:fractional_matter}, \eqref{eq:continuity_final} and \eqref{eq:omega_expression} into the field equations gives 
\begin{align}
    \label{eq:W00_sub}
    \dut{W}{0}{0}: &&  0 &= \mathcal{F}_{1}(\varphi, \psi, \dot{\psi}, \delta_{m}, \dot{\delta}_{m})\,, \\
    \label{eq:W0i_sub}
    \dut{W}{0}{i}: &&  0 &= \mathcal{F}_{2}(\varphi, \psi, \dot{\psi}, \dot{\delta}_{m})\,, \\
    \label{eq:Wi0_sub}
    \dut{W}{i}{0}: &&  0 &= \mathcal{F}_{3}(\varphi, \psi, \dot{\psi}, \dot{\delta}_{m})\,, \\
    \label{eq:Wii_sub}
    \dut{W}{i}{i}: &&  0 &= \mathcal{F}_{4}(\varphi, \dot{\varphi}, \psi, \dot{\psi}, \ddot{\psi})
\end{align}
where $\mathcal{F}_{i}$ are functions defined in Appendix~\ref{app:FE_functions}. Eqs~\eqref{eq:W0i_sub} [$\dut{W}{0}{i}$] and \eqref{eq:Wi0_sub} [$\dut{W}{i}{0}$] are simultaneously solved to obtain equations for $\varphi$ and $\psi$ in terms of higher derivatives of the form
\begin{align}
    \label{eq:varphi_expression}
    \varphi &= \mathcal{G}_{1}(\dot{\psi},\dot{\delta}_{m})\,, \\
    \label{eq:psi_expression}
    \psi &= \mathcal{G}_{2} (\dot{\psi},\dot{\delta}_{m})\,,
\end{align}
where $\mathcal{G}_{i}$ are functions defined in Appendix~\ref{app:FE_functions}. Further $\mathcal{F}_{i}$ and $\mathcal{G}_{i}$ functions are not included due to the complicated nature of the expressions. By differentiating both Eqs.~\eqref{eq:varphi_expression} and \eqref{eq:psi_expression}, new equations of motions are constructed in the form
\begin{align}
    \label{eq:DW0i}
    0 &= \mathcal{F}_{5}(\dot{\varphi}, \dot{\psi}, \ddot{\psi}, \dot{\delta}_{m}, \ddot{\delta}_{m}) \,, \\
    \label{eq:DWi0}
    0 &= \mathcal{F}_{6}(\dot{\psi}, \ddot{\psi}, \dot{\delta}_{m}, \ddot{\delta}_{m})  \,,
\end{align}
such that the zeroth order derivatives do not appear. Once again, this set of equations is solved simultaneously, this time for $\dot{\varphi}$ and $\dot{\psi}$ in terms of higher order derivatives, $\delta_{m}$ and its derivatives:
\begin{align}
    \label{eq:Dvarphi_expression}
    \dot{\varphi} &= \mathcal{G}_{3}(\ddot{\psi}, \dot{\delta}_{m},\ddot{\delta}_{m}) \,,\\
    \label{eq:Dpsi_expression}
    \dot{\psi} &= \mathcal{G}_{4}(\ddot{\psi}, \dot{\delta}_{m}, \ddot{\delta}_{m})\,.
\end{align}
From  all  equations in this section, it is clear that the mode $\ddot{\varphi}$ does not appear anywhere in the system. Hence, we will only obtain an expression for the $\ddot{\psi}$ mode. Consider Eq.~\eqref{eq:W00_sub} [$\dut{W}{0}{0}$], which we differentiate with respect to the cosmic time in order to obtain an equation of the form
\begin{align}
    \label{eq:DW00}
    0 &= \mathcal{F}_{7}(\varphi, \dot{\varphi},  \psi, \dot{\psi}, \ddot{\psi}, \delta_{m}, \dot{\delta}_{m}, \ddot{\delta}_{m})\,.
\end{align}
By substituting the equations for zeroth and first order derivatives (\ref{eq:varphi_expression}, \ref{eq:psi_expression}, \ref{eq:Dvarphi_expression}, \ref{eq:Dpsi_expression}) given in terms of $\mathcal{G}_{i}$ functions, an equation for $\ddot{\psi}$ in terms of $\delta_{m}$ and its derivatives is obtained
\begin{align}
    \label{eq:DDpsi_expression}
    \ddot{\psi} = \mathcal{G}_{5}(\delta_{m}, \dot{\delta}_{m}, \ddot{\delta}_{m})\,.
\end{align}
Next, the M\'{e}sz\'{a}ros equation can be derived directly from the continuity equation~\eqref{eq:continuity_2} and differentiating with respect to the time. Substituting $\mathcal{G}_i$ functions in increasing order of $i$ results in
\begin{align}
    \label{eq:meszaros_eq_in_t}
    0 &= \ddot{\delta}_{m} + 2 H\,\mathcal{H}_{1} \dot{\delta}_{m} - \frac{\kappa^{2} \rho}{2} \mathcal{H}_{2} \delta_{m}\,,
\end{align}
where
\begin{align}
    \label{eq:coeff_Ddeltam}
    \mathcal{H}_{1} &= 1 + 
    \Bigg(\frac{18 \frac{a^{2}}{k^{2}}}{1 + f_{T} - 36 \frac{a^{2}}{k^{2}} H^{2} \dot{H} f_{TT}} \Bigg) \Bigg(  (2 \dot{H}^{2} + H \ddot{H} + 2 H^{2} \dot{H}) f_{TT} - 12 H^{2} \dot{H}^{2} f_{TTT} 
   + \frac{6 H^{2} \dot{H}^{2} f_{TT}^{2}}{1+f_{T}} \Bigg) \,, \\ \nonumber \\
    \label{eq:coeff_deltam}
    \mathcal{H}_{2} &= \frac{1}{1 + f_{T} - 36 \frac{a^{2}}{k^{2}} H^{2} \dot{H} f_{TT}} 
    - \frac{36 \frac{a^{2}}{k^{2}}}{(1 + f_{T}) (1 + f_{T} - 36 \frac{a^{2}}{k^{2}} H^{2} \dot{H} f_{TT})} \Bigg( f_{TT} (3 H^{2} \dot{H} + 2 \dot{H}^{2} + H \ddot{H}) \nonumber \\
    & \qquad  12 H^{2} \dot{H}^{2} \left(- f_{TTT}  + \frac{ f_{TT}^{2} }{1+f_{T}}\right) \Bigg)\,.
\end{align}
The M\'{e}sz\'{a}ros equation can be simplified by rewriting the well-known expressions
\begin{align}
    \label{eq:Hdot_Hddot}
    \dot{H} = \kappa_{1} H^{2}\,, \qquad \text{and} \qquad \ddot{H} = \kappa_{2} H^{3}\,,
\end{align}
where $\kappa_{1} = -\frac{3}{2}$ and $\kappa_{2} = \frac{9}{2}$ for the matter dominated era where $a(t) \propto t^{\frac{2}{3}}$~\cite{delaCruz-Dombriz:2008ium}. At this point, the subhorizon limit $k \gg aH $ can be applied. By rewriting Eq.~\eqref{eq:meszaros_eq_in_t} using $\xi = \frac{aH}{k}$ and performing a Taylor expansion of $\xi$ about 0 up to first order, it follows that
\begin{align}
     \label{eq:coeff_Ddeltam_sub}
    \mathcal{H}_{1}^{\text{sub}} &= 1 + 
    \Bigg(\frac{18 \xi^{2} H^{2}}{1 + f_{T} - 36 \xi^{2} \kappa_{1} H^{2} f_{TT}} \Bigg) \Bigg(  (2 \kappa_{1}^{2} + \kappa_{2}  + 2 \kappa_{1} ) f_{TT} - 12 \kappa_{1}^{2} H^{2} f_{TTT} 
   + \frac{6 \kappa_{1}^{2} H^{2}  f_{TT}^{2}}{1+f_{T}} \Bigg) \nonumber \\
   & \rightarrow  1  \, \\
    \label{eq:coeff_deltam_sub}
    \mathcal{H}_{2}^{\text{sub}} &= \frac{1}{1 + f_{T} - 36 \xi \kappa_{1} H^{2} f_{TT}} 
    - \frac{36 \xi^{2} H^{2}}{(1 + f_{T}) (1 + f_{T} - 36 \xi^{2} \kappa_{1} H^{2} f_{TT})} \Bigg( f_{TT} (3 \kappa_{1} + 2 \kappa_{1}^{2} + \kappa_{2} ) \nonumber \\
    & \qquad  12 \kappa_{1}^{2} H^{2} \left(- f_{TTT}  + \frac{ f_{TT}^{2} }{1+f_{T}}\right) \Bigg)  \rightarrow \frac{1}{1+f_{T}}\,,
\end{align}
which is the result obtained in Ref.~\cite{Zheng:2010am}. Additionally, as the Universe becomes dominated by dark energy, we should consider the de Sitter phase where $a(t) \propto e^{H_{\text{ds}}t}$ where $H_{\text{dS}}$ is a positive constant. Thus, Eq.~(\ref{eq:coeff_Ddeltam}-\ref{eq:coeff_deltam}) reduce to
\begin{align}
    \mathcal{H}_{1}^{\text{dS}} & = 1\,,\\
    \mathcal{H}_{2}^{\text{dS}} &= \frac{1}{1+f_{T}}\,,
\end{align}
the same result as the subhorizon limit. This allows us to consider the general case of Eq.~\eqref{eq:meszaros_eq_in_t} running from  the matter epoch up to the current times, where dark energy dominates the energy budget of the Universe. However, it should be noted that since $k$ is typically expressed as a product of the Hubble constant $H_{0}$, certain terms in Eqs~(\ref{eq:coeff_Ddeltam}-\ref{eq:coeff_deltam}) could restrict the $k$ value for which the subhorizon limit in Eqs~(\ref{eq:coeff_Ddeltam_sub}-\ref{eq:coeff_deltam_sub}) can be obtained. Hence, we will proceed with the general M\'{e}sz\'{a}ros equation in the case where the subhorizon limit is exceeded or $k$-dependencies persist. 

\subsection{The growth factor} \label{subsec:growth_factor}

Growth is defined as the ratio of perturbation amplitude at scale factor $a$ to its value at some initial scale factor $a_{i}$
\begin{align}
    D(a) = \frac{\delta_{m}(a)}{\delta_{m}(a_{i})}\,.
\end{align}
Hence, the  M\'{e}sz\'{a}ros equation~\eqref{eq:meszaros_eq_in_t} is expressed in terms of the scale factor 
\begin{align}
    \label{eq:meszaros_in_a}
    \delta_{m}''(a) + \left(\frac{1 + 2 \mathcal{H}_{1}^{(a)}}{a} + \frac{h'(a)}{h(a)}\right) \delta_{m}'(a) - \frac{3}{2} \Omega_{m} a^{-3} \mathcal{H}_{2}^{(a)} \delta_{m}(a) = 0\,,
\end{align}
where primes (') denote derivatives with respect to $a$ and the energy density and matter density parameter are given by, respectively,
\begin{align}
    \rho(t) &= \frac{3 H_0^2 \Omega_{m0}}{\kappa^2} a^{-3}\,, \\ 
    \label{eq:omega_m}
    \Omega_{m}(a) &= \frac{\Omega_{m0} a^{-3}}{h^{2}(a)}\,,
\end{align}
where $h(a) = \frac{H(a)}{H_{0}}$ is the normalized Hubble parameter, $\mathcal{H}_{1}^{(a)}$ and $\mathcal{H}_{2}^{(a)}$ are the expressions given by Eqs~\eqref{eq:coeff_Ddeltam} and \eqref{eq:coeff_deltam} transformed in terms of $a$, and $\Omega_{m0}$ is the density parameter at current time $t_{0}$ where $a(t_{0}) = 1$.
The initial conditions for the second order differential equation are taken to be $D(a_{i}) = 1$ and $D'(a_{i}) = 10$~\cite{Dodelson:2003ft}, where $a_{i} = 0.1$ in order to ensure that the system is describing the matter dominated era. The models considered in the next section do not result in an analytical solution, hence a numerical approach is considered requiring a numerical solution $h(a)$ for each separate model. Note, the analysis will be carried out for both the general case dependent on a variety of $k$ values and the subhorizon limit investigated in Ref.~\cite{Zheng:2010am} to highlight differences or similarities between the two approaches.

\subsection{The growth index} \label{subsec:growth_index}

In this section, we consider the scenario where the growth index $\gamma$ varies with redshift. Starting from the M\'{e}sz\'{a}ros equation~\eqref{eq:meszaros_in_a}, the growth rate of clustering is given by~\cite{peebles1993principles}
\begin{align} \label{eq:g_func}
    g(a) = \frac{d \ln \delta_{m}}{d \ln a} \simeq \Omega_{m}(a)^{\gamma(a)}\,,    
\end{align}
wherein the growth factor can be rewritten as
\begin{align}
    D(a) = \exp \left( \int_{1}^{a} \frac{\Omega_{m}(\tilde{a})^{\gamma(\tilde{a})}}{\tilde{a}} d\tilde{a} \right)\,,
\end{align}
We will preemptively define the derivative of Eq.~\eqref{eq:omega_m}
\begin{align}
    \label{eq:Domega_m}
    \Omega_{m}'(a) &= - 3 \frac{\Omega_{m}(a)}{a} \left(1 + \frac{2}{3} \frac{a h'(a)}{h(a)}\right)\,.
\end{align}
By substituting Eq.~\eqref{eq:g_func}, along with Eq.~\eqref{eq:Domega_m}, into the  M\'{e}sz\'{a}ros equation~\eqref{eq:meszaros_in_a} to obtain
\begin{align}
    \frac{3}{2} \mathcal{H}_{2}^{(a)} \Omega_{m}(a) = \Omega_{m}(a)^{\gamma(a)} \left( \Omega_{m}(a)^{\gamma(a)} + a (1 - 2 \gamma(a)) \frac{h'(a)}{h(a)} + 2 \mathcal{H}_{1}^{(a)} - 3 \gamma(a) + a \ln \Omega_{m} \gamma'(a) \right)\,. 
\end{align}
Evaluating the above equation at current time gives
\begin{align}
    \frac{3}{2} \mathcal{H}_{2}^{(a=1)} \Omega_{m0} = \Omega_{m0}^{\gamma(1)} \left( \Omega_{m0}^{\gamma(1)} + (1 - 2 \gamma(1)) h'(1) + 2 \mathcal{H}_{1}^{(a = 1)} - 3 \gamma(1) + \ln \Omega_{m0} \gamma'(1) \right)\,.
\end{align}
A particular parametrization of dynamical growth index is given by $\gamma(a) = \gamma_0 + \gamma_1 (1-a)$~\cite{Polarski:2007rr,Pouri:2014nta,Dossett:2010gq,Basilakos:2016xob}. Here, we will adopt the constant growth index case by setting $\gamma_{1} = 0$, simplifying the previous equation to
\begin{align} \label{eq:growth_index_differential}
    0  =  \frac{\Omega_{m0}^{\gamma} + (1 - 2 \gamma) h'(1) + 2 \mathcal{H}_{1}^{(a = 1)} - 3 \gamma - \frac{3}{2} \mathcal{H}_{2}^{(a=1)} \Omega_{m0}^{1-\gamma}}{\ln \Omega_{m0}} \,,
\end{align}
where $\gamma_{0} = \gamma$ from this point forward. Once again, the solution for $h'(1)$ would need to be obtained on a case-by-case basis.

\section{Growth in different \texorpdfstring{$f(T)$}{} models}\label{sec:models}

Here, we will consider the following five $f(T)$ models: Power Law~(\ref{subsec:power_law}), Linder~(\ref{subsec:f2_model}), Exponential~(\ref{subsec:exponential}), Logarithmic~(\ref{subsec:log}) and Hyperbolic-Tangent~(\ref{subsec:tanh})~\cite{Nesseris:2013jea}. While the first 3 models can attain the $\Lambda$CDM limit, it is not the case for the Logarithmic and Hyperbolic-Tangent models. However, they could result interesting from a fundamental physics point of view as interesting effective field models \cite{Cai:2015emx, Dvali:2000hr}. Regardless, a constant model $f(T) = c$, where $c$ is the constant, is taken into consideration for all cases. Additionally, the $\Lambda$CDM results, obtained from Planck collaboration observations, will be included for comparison i.e. $\Omega_{m0} = 0.315$ and $H_{0} = 67.4\, \text{km}\,\text{s}^{-1}\,\text{Mpc}^{-1}$~\cite{Planck:2018vyg}. The constant growth index value for these values results in 
\begin{align} \label{eq:gamma_lambdaCDM_Planck}
    \gamma_{\text{$\Lambda$CDM}} = 0.5595\,.
\end{align}

The parameters $H_{0}$, $\Omega_{m0}$ and $\beta_{i}$ constant ($i \in [1,5]$ corresponding to different models) for each model are obtained from different combination of observational data sets. Here we will use the results presented in Ref.~\cite{Briffa:2021nxg}. These results are based on the following assumptions:
\begin{itemize}
    \item Cosmic Chronometers (CC): Massive passively evolving galaxies, which build their mass at high redshift over a short period of time and exhaust gas reservoir at early stages, can be used as CC. By considering two of such galaxies with a small redshift difference, the age difference can be determined~\cite{Jimenez:2001gg}. The compilation of data sets~\cite{Jimenez:2003iv, Zhang:2012mp, Moresco:2016mzx, Simon:2004tf, Moresco:2012jh, Stern:2009ep, Moresco:2015cya} considers different ranges of redshift $z$ up to $z \lesssim 2$; some of them adopts the differential age method and others analyse the differential D4000 breaking which requires stellar population synthesis models for calibration.
    \item Supernovae Type Ia (SNeIa): These supernovae, characterised by the absence of hydrogen lines in spectra, yield to exceptional brightness and a uniform peak luminosity, making them ideal standard candles to measure distances in the Universe. In particular, the Pantheon compilation of SNeIa (SN), using multiple surveys measuring relative luminosity distances in the redshift range $0.01 < z < 2.26$~\cite{Pan-STARRS1:2017jku}, is utilized.
    \item Baryonic Acoustic Oscillations (BAO): Remnants of sound waves which propagated through the early universe leave an imprint on the large-scale structure, creating characteristic peaks in galaxy clustering at the sound horizon scale. BAO data sets make use of the Sloan Digital Sky Survey (SDSS) Data Set 7~\cite{Ross:2014qpa}, six degree Field Galaxy Survey (6dFGS)~\cite{Beutler:2011hx}, Ly$\alpha$-forest flux transmission using complete SDSS-III~\cite{duMasdesBourboux:2017mrl}, the Baryon Oscillation Spectroscopic Survey (BOSS)~\cite{BOSS:2016wmc} and the extended BOSS Data Release 14 quasar (DR14Q) survey~\cite{Zhao:2018gvb}. 
\end{itemize}
The $H_{0}$ priors come from:
\begin{itemize}
    \item SH0ES: Estimates of $H_{0}$ (R19) from the observation of 70 long-period Cepheids in the Large Magellanic Cloud (LMC) from the Hubble Space Telescope (HST)~\cite{Riess:2019cxk}. 
    \item H0LiCOW Collaboration: Measurements of $H_{0}$ using gravitational lensing by determining time-delay distance $D_{\Delta t}$, which is sensitive to $H_{0}$~\cite{Wong:2019kwg}.
    \item Tip of the Red Giant Branch (TRGB): TRGB occurs as stars on the red giant branch experiences helium burning, hence starts transitioning to the horizontal branch as luminosity decrease. At the TRGB point, the stars are bright and red, providing an independent framework to determine $H_{0}$~\cite{Freedman:2019jwv}.
\end{itemize}

For all models, the rest of the parameters are obtained from background equations in order to yield to the normalised background first Friedman equation~\eqref{eq:background_friedmann_1}, and using the constrained values of the rest of parameters as determined by the combination of different data sets. The constant growth index is determined using~Eq.~\eqref{eq:growth_index_differential} for which $\gamma_{1} = 0$ for the subhorizon limit.  The result is obtained by the M\'{e}sz\'{a}ros equation with coefficients given by Eqs~(\ref{eq:coeff_Ddeltam_sub}-\ref{eq:coeff_deltam_sub}), following Ref.~\cite{Zheng:2010am} results. It should be noted that, from this point on, $\gamma_{0} = \gamma$. Next, we use the general form with $k$-dependencies, given by Eqs~(\ref{eq:coeff_Ddeltam}-\ref{eq:coeff_deltam}), to determine at which value of $k$ does the subhorizon limit come into play in order to retrieve a value for $\gamma$ identical up to 4 decimal places. Additionally, the $k$-cutoff at radiation-matter equality, $k_{\text{eq}}$, is determined for comparison. Hence, we are able to numerically solve Eq.~\eqref{eq:meszaros_in_a} to plot the growth factor $D(a)$ for different values of $k$. In all data sets, $H_{0}$ is expressed in terms of $\text{km}\,\text{s}^{-1}\,\text{Mpc}^{-1}$ units.

\subsection{\texorpdfstring{$f_1(T)$}{}: Power Law Model}\label{subsec:power_law}

First, we considered the Power Law model given by~\cite{Bengochea:2008gz} , which is promising in producing important elements of late-time accelerated cosmic expansion, through
\begin{align}
    f_1(T) = -\alpha_{1}(-T)^{\beta_{1}}\,,
\end{align}
where $\alpha_{1}$ and $\beta_{1}$ are constants. The $\Lambda$CDM limit can be obtained for $\beta = 0$, and $\beta 
 = \tfrac{1}{2}$ results in a Dvali, Gabadadze and Porrati (DGP) model~\cite{Dvali:2000hr} term. The parameter $\alpha_{1}$ can be determined at any  time, hence the first Friedman equation~\eqref{eq:background_friedmann_1} is evaluated at current time $t_{0}$ such that $a(t_{0}) = 1$ and $H(t_{0}) = H_{0}$, which gives
\begin{align}
    \alpha_{1} = \frac{(6 H_{0}^2)^{1 - \beta_{1}} (1 - \Omega_{m0})}{ 2 \beta_{1}-1}\,,
\end{align}
such that Eq.~\eqref{eq:background_friedmann_1} can be rewritten as
\begin{align}
    h(a)^{2} = \Omega_{m0} a^{-3} + (1 - \Omega_{m0}) h(a)^{2\beta_{1}}\,,
\end{align}
where $h(a) = \frac{H(a)}{H_{0}}$ and $h(1) = H_{0}$. Observations at current time point out  that the Power Law model fits for $\beta_1 \ll 1$, resulting in $f_{1,T} \ll 1$ during the matter era~\cite{Linder:2010py}. Hence, further derivatives of function $f_{1}(T)$ are even smaller.

In the Power Law model, the $k$-dependency comes into play when looking into higher index precision values. If we consider the constant $\gamma$ limit of Eq.~\eqref{eq:growth_index_differential}, Table~\ref{tab:power_law} provides a list of $\gamma$ values which result from different data sets when taking the subhorizon limit, some with $H_{0}$ priors. From these results, the values of $k$ at which the $\gamma$ corresponds to that obtained from the subhorizon limit, up to 4 decimal places (d.p.), can be determined. For the Power Law model, it is important to consider where the subhorizon condition $k \gg a H$ starts to apply, as not all $k$ values, larger than the matter-radiation equality $k_{\text{eq}}$ value, necessarily are within the subhorizon regime. The largest $\gamma$ index is given by CC+SN-BAO set at $0.5718$, while the lowest value is for CC+SN with $0.5284$. At face-value, the growth index results seem to correlate with the $\beta_{1}$ value of the model, as small model indices yield smaller $\gamma$ values. Additionally, CC+SN+R19 set has the higher value of $k$ for which the subhorizon results matches up to 4~d.p. 

\begin{table}[h!]
\footnotesize
    \centering
    \begin{tabular}{|c|>{\centering\arraybackslash}p{1cm}|>{\centering\arraybackslash}p{1cm}|>{\centering\arraybackslash}p{1cm}|>{\centering\arraybackslash}p{1.8cm}|>{\centering\arraybackslash}p{1.8cm}|>{\centering\arraybackslash}p{1.8cm}|}
    \hline
    \multicolumn{7}{|c|}{Power Law Model}\\ \hline
        Data Set & $\beta_1$ & $H_{0}$ & $\Omega_{m0}$ & $k_{\text{eq}} / H_{0}$ & $k/H_{0}$ & $\gamma$
        \\ \hline\hline
        CC + SN & $-0.22$ & $68.5$ & $0.35$ & $54.0062$ & $ > 411.89$ & $0.5284$ 
        \\ 
        CC + SN + R19 & $-0.13$ & $71.3$ & $0.326$  & $50.3029$ & $>1130.40$ & $0.5398$
        \\
        CC + SN + HW & $-0.16$ & $71.0$ & $0.329$ & $50.7658$ & $>1422.26$ & $0.5361$
        \\
        CC + SN + TRGB & $-0.20$ & $69.1$ & $0.344$ & $53.0803$ & $>327.57$ &  $0.5308$
        \\
         CC + SN + BAO & $0.06$ & $67.1$ & $0.294$ & $45.3652$  & $>204.17$ & $0.5718$ \\

        CC + SN + BAO + R19 & $-0.14$ & $69.9$ & $0.305$ &  $47.0625$ & $>176.48$ & $0.5400$
        \\
        CC + SN + BAO + HW & $-0.12$ & $67.7$ & $0.304$ &  $46.9082$ & $>565.63$ & $0.5424$
        \\
        CC + SN + BAO + TRGB & $-0.01$ & $68.1$ & $0.298$ & $45.9824$ & $>123.77$ & $0.5585$
        \\\hline
    \end{tabular}
    \caption{Power Law model results for different data sets constraints. The second column corresponds to the $\beta_{1}$ parameter in the models, the third column corresponds to the $H_{0}$ value and the fourth column gives the $\Omega_{m0}$ value. For each set, the $k$-cutoff value at matter-radiation equality is determined as $k_{\text{eq}}$ in the fourth column. The fifth column gives the minimum value for $k$ in order for the M\'{e}sz\'{a}ros equation to start behaving like the subhorizon limit, particularly determined by the constant $\gamma$ value in the sixth column up to 4~d.p.}
    \label{tab:power_law}
\end{table}

\begin{figure}[h!]
    \subfloat[CC + SN]{\includegraphics[width=0.3\textwidth]{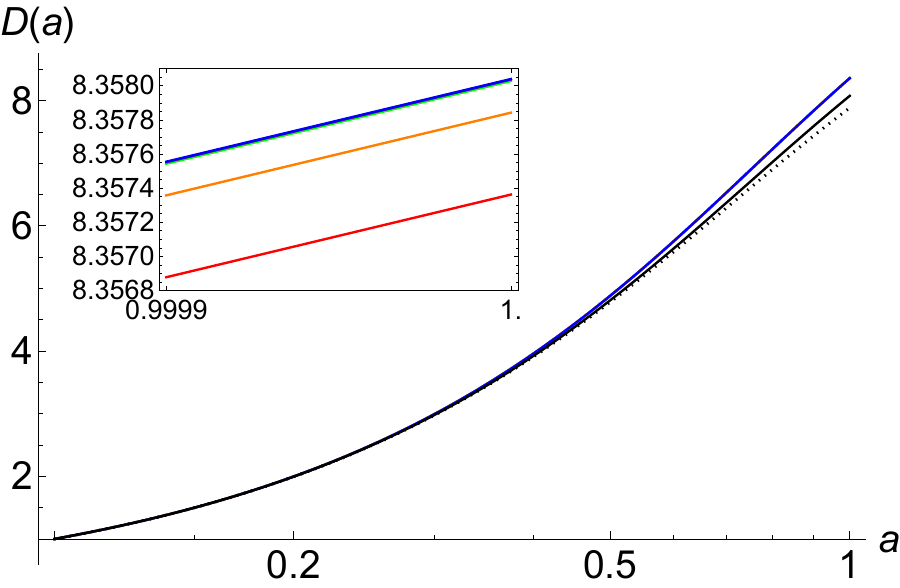}
    \label{fig:powerlaw_CC_SN}}
    \hfill
    \subfloat[CC + SN + BAO]
    {\includegraphics[width=0.3\textwidth]{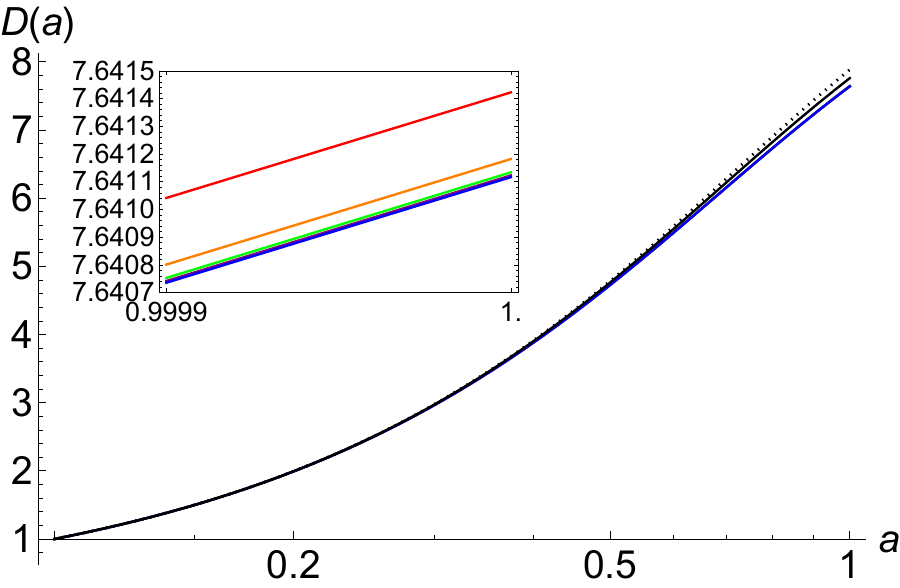}
    \label{fig:powerlaw_CC_SN_BAO}}
    \hfill
    \subfloat[CC + SN + R19]
    {\includegraphics[width=0.3\textwidth]{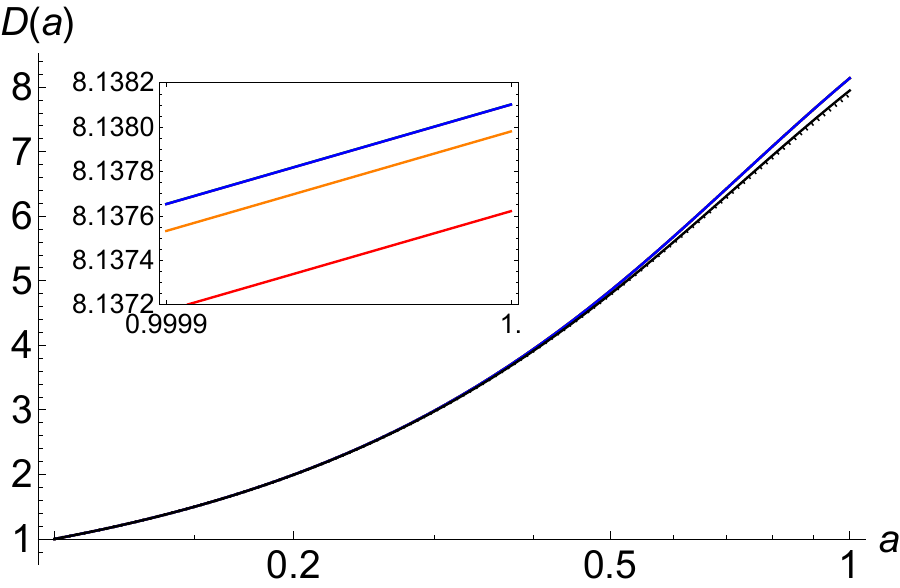}
    \label{fig:powerlaw_CC_SN_R19}}
    \vskip 0.1cm
    \subfloat[CC + SN + HW]
    {\includegraphics[width=0.3\textwidth]{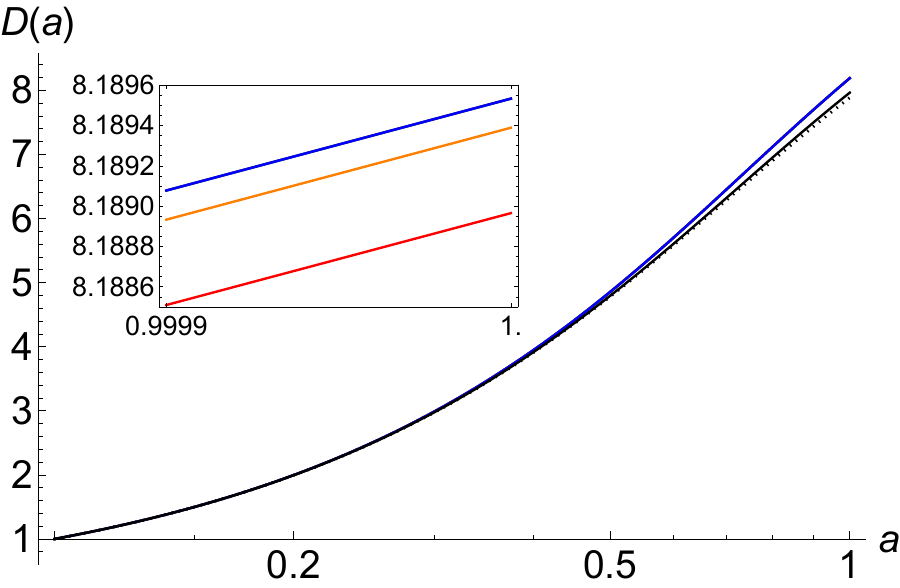}
    \label{fig:powerlaw_CC_SN_HW}}
    \hfill
    \subfloat[CC + SN + TRGB]
    {\includegraphics[width=0.3\textwidth]{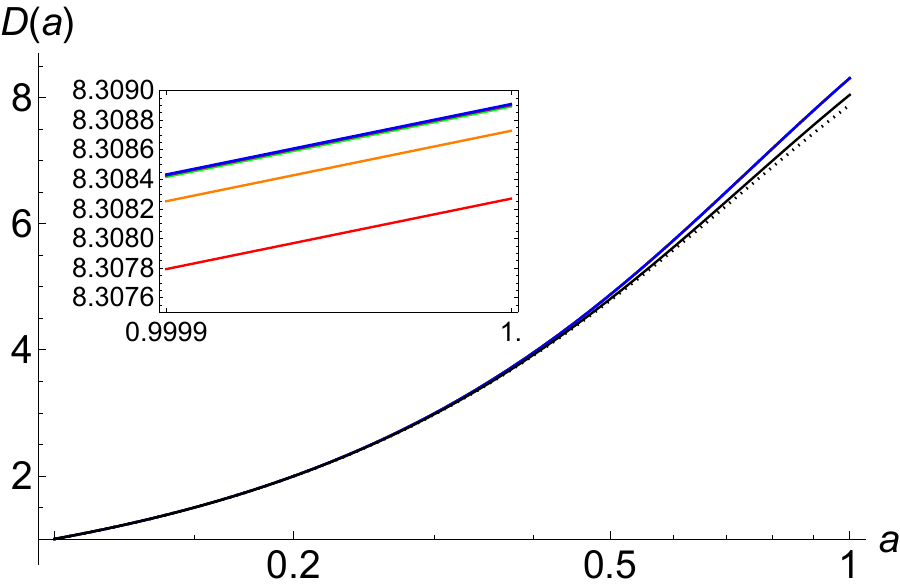}
    \label{fig:powerlaw_CC_SN_TRGB}}
    \hfill
    \subfloat[CC + SN + BAO + R19]
    {\includegraphics[width=0.3\textwidth]{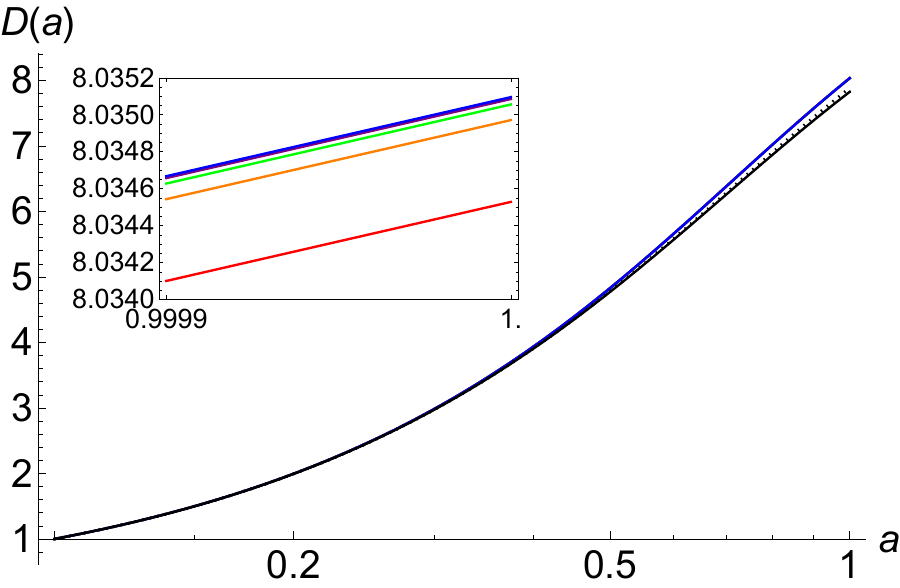}
    \label{fig:powerlaw_CC_SN_BAO_R19}}
    \vskip 0.1cm
    \subfloat[CC + SN + BAO + HW]
    {\includegraphics[width=0.3\textwidth]{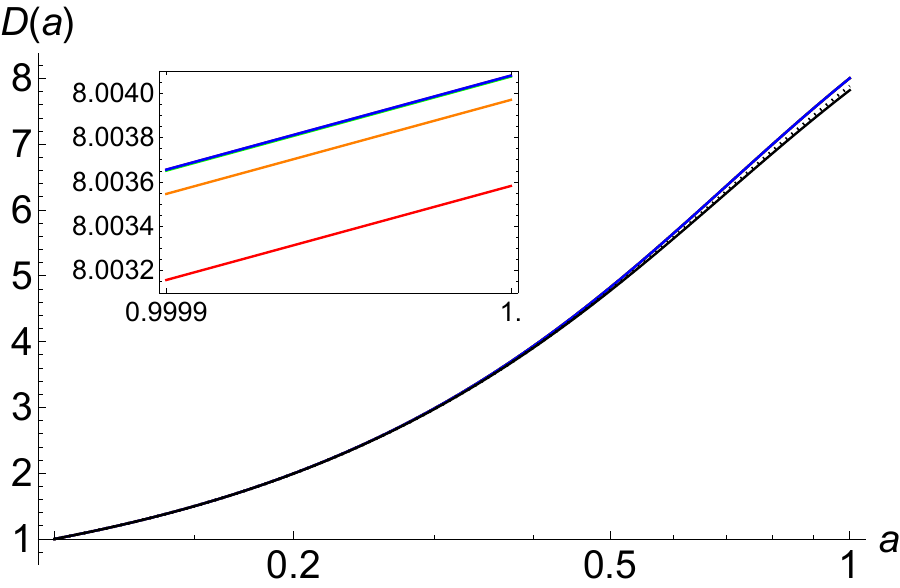}
    \label{fig:powerlaw_CC_SN_BAO_HW}}
    \hfill
    \subfloat[CC + SN + BAO + TRGB]
    {\includegraphics[width=0.3\textwidth]{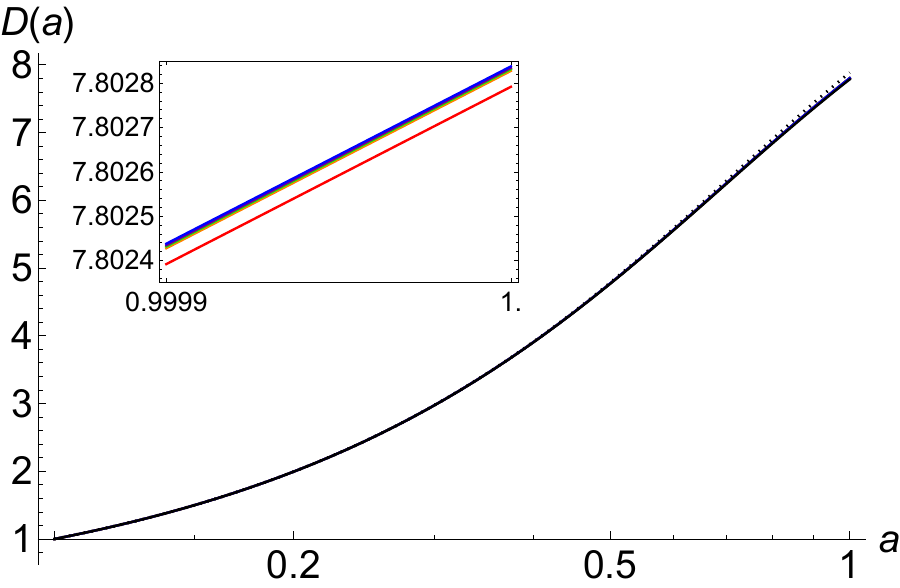}
    \label{fig:powerlaw_CC_SN_BAO_TRGB}}
    \hfill
    \hspace{0.06\textwidth}
    \subfloat
    {\includegraphics[width=0.18\textwidth]{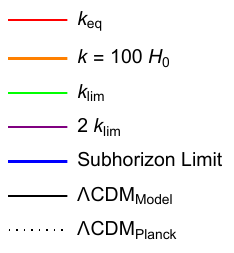}}
    \hspace{0.06\textwidth}
    \caption{Growth factor plots for the Power Law model. Red line corresponds to the growth factor solution at radiation-matter equality, the green one represents the solution at the point just before the subhorizon limit applies ($k_{\text{lim}}$), and the blue line is the subhorizon solution. Black, solid line is the growth factor at $\Lambda$CDM limit obtained for the model as $\beta_{1} \rightarrow 0$ ($\Lambda\text{CDM}_{\text{Model}}$), and the dotted line is the $\Lambda$CDM solution for the results obtained in the Planck collaboration ($\Lambda\text{CDM}_{\text{Planck}}$)~\cite{Planck:2018vyg}. The yellow curve, where $k = 2 k_{\text{lim}}$, is not visible in the scales presented here as these values are well within the subhorizon limit and can be depicted by the blue curve.}
    \label{fig:powerlaw}
\end{figure}

The changes with $k$ have been further highlighted in Figs.~\ref{fig:powerlaw} when solving for the growth factor $D(a)$. In all cases, the $k_{\text{eq}}$  and $k = 100 H_{0}$ values are both further away from the subhorizon limit as these are below the required regime. But, for the value given in the sixth column of Table~\ref{tab:power_law}, given the green curve in the upper right quadrant of each plot, it can be seen approaching the subhorizon limit, and any value above the $k$ value (yellow line) coincides with the subhorizon curve and hence, not visible. It should be noted that such discrepancies are mainly seen when considering 5 significant figure changes (upper-right quadrants), while when considering the growth at $0.1 \leq a \leq 1$ the subhorizon limit considered in Ref.~\cite{Zheng:2010am} suffices. Finally, for the majority of the sets considered, the growth factor increases more than its corresponding $\Lambda$CDM limit, except for the case of CC+SN+BAO~(\ref{fig:powerlaw_CC_SN_BAO}). This is mainly due to the positive nature of $\beta_{1}$ in the CC+SN+BAO constraints.

\subsection{\texorpdfstring{$f_2(T)$}{}: Linder Model} \label{subsec:f2_model}

As for the Power-Law model setting, the Linder model~\cite{Linder:2010py} was originally designed in the context of late-time cosmological concerns but it has  shown to be more widely applicable. It is given by
\begin{align}
    f_{2}(T) = -\alpha_{2} T_{0} \left( 1 - \exp{\left(- \beta_{2} \sqrt{\frac{T}{T_{0}}}\right)} \right)\,,
\end{align}
where $T_0 = -6{H_0}^2$, and 
\begin{align}
    \alpha_{2} = \frac{e^{\beta_{2}} (1 - \Omega_{m0})}{e^{\beta_{2}}-1-\beta_{2}}\,,
\end{align}
and $\beta_{2}$ are constants such that Friedmann equation~\eqref{eq:background_friedmann_1} simplifies to
\begin{align}
    h(a)^{2} = \Omega_{m0} a^{-3} + (1 - \Omega_{m0}) \frac{1 - e^{-\beta_{2}h(a)} (1 + \beta_{2} h(a))}{1 -e^{-\beta_{2}}(1+\beta_{2})}\,,
\end{align}
such that the $\Lambda$CDM limit can be obtained for $\beta_{2} \rightarrow +\infty$. 

\begin{figure}[ht!]
    \centering
    \subfloat[CC + SN]
    {\includegraphics[width=0.3\textwidth]{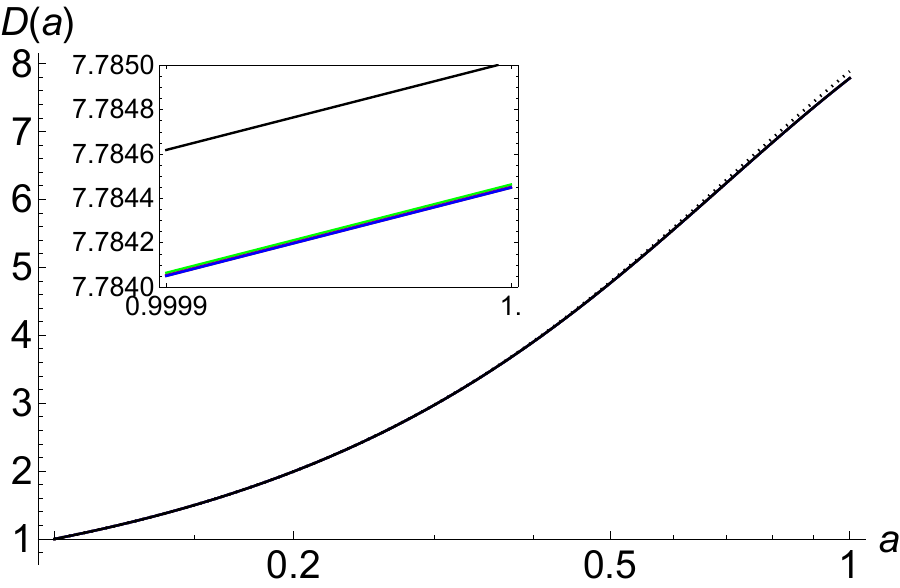}
    \label{fig:linder_CC_SN}}
    \hfill
    \subfloat[CC + SN + R19]
    {\includegraphics[width=0.3\textwidth]{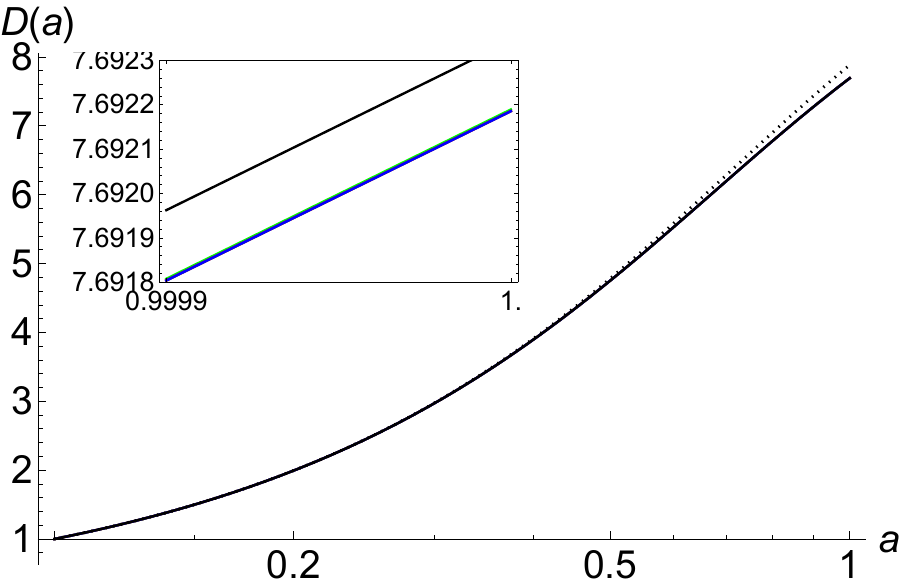}
    \label{fig:linder_CC_SN_R19}}
    \hfill
    \subfloat[CC + SN + HW]
    {\includegraphics[width=0.3\textwidth]{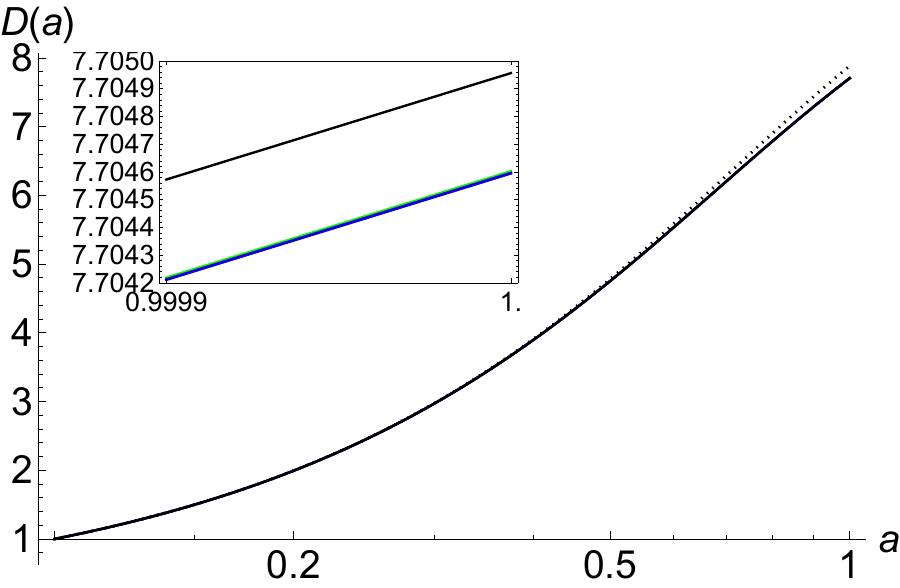}
    \label{fig:linder_CC_SN_HW}}
    \vskip 0.1cm
    \subfloat[CC + SN + TRGB]
    {\includegraphics[width=0.3\textwidth]{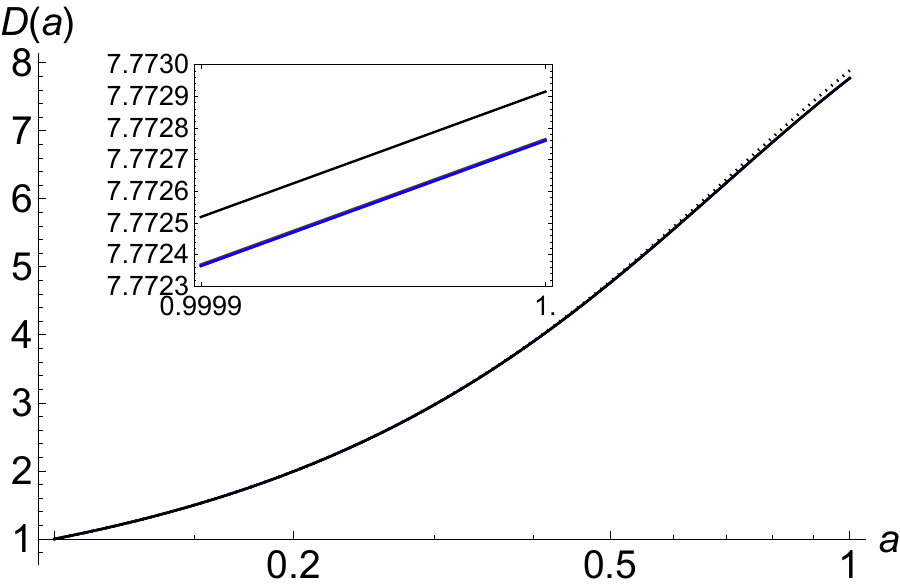}
    \label{fig:linder_CC_SN_TRGB}}
    \hfill
    \subfloat[CC + SN + BAO]
    {\includegraphics[width=0.3\textwidth]{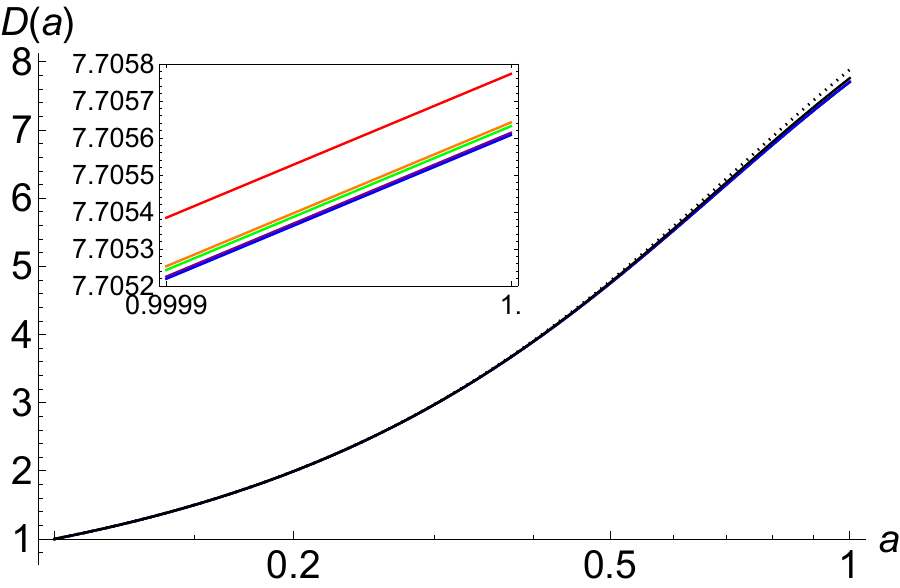}
    \label{fig:linder_CC_SN_BAO}}
    \hfill
    \subfloat[CC + SN + BAO + R19]
    {\includegraphics[width=0.3\textwidth]{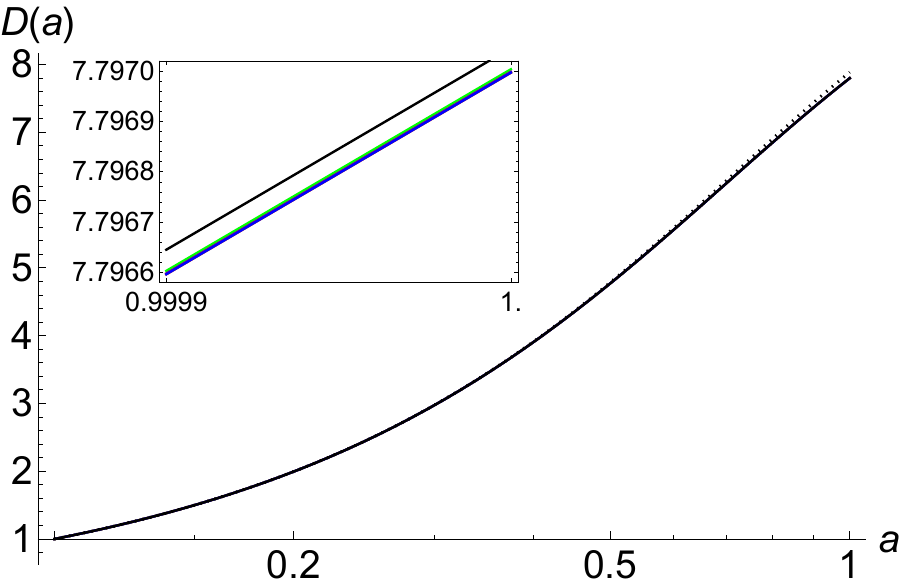}
    \label{fig:linder_CC_SN_BAO_R19}}
    \vskip 0.1cm
    \subfloat[CC + SN + BAO + HW]
    {\includegraphics[width=0.3\textwidth]{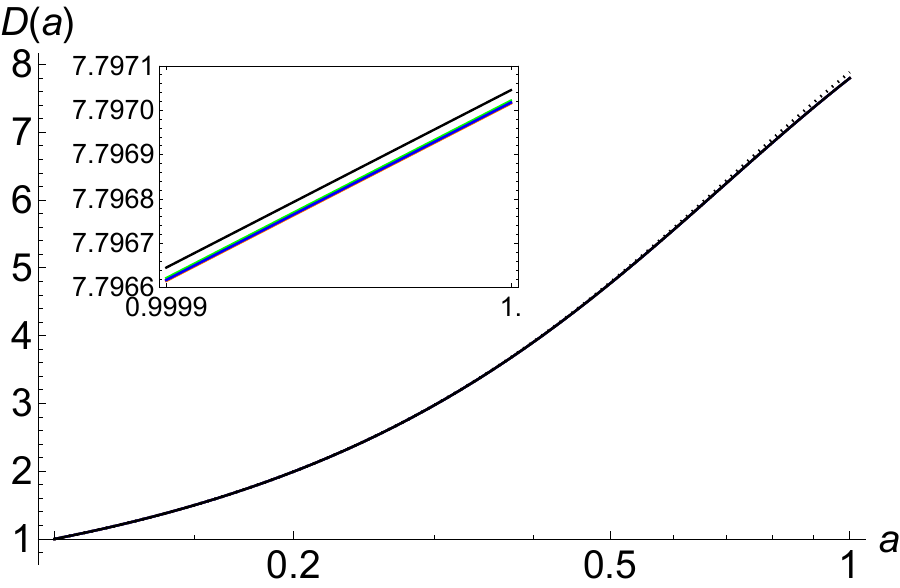}
    \label{fig:linder_CC_SN_BAO_HW}}
    \hfill
    \subfloat[CC + SN + BAO + TRGB]
    {\includegraphics[width=0.3\textwidth]{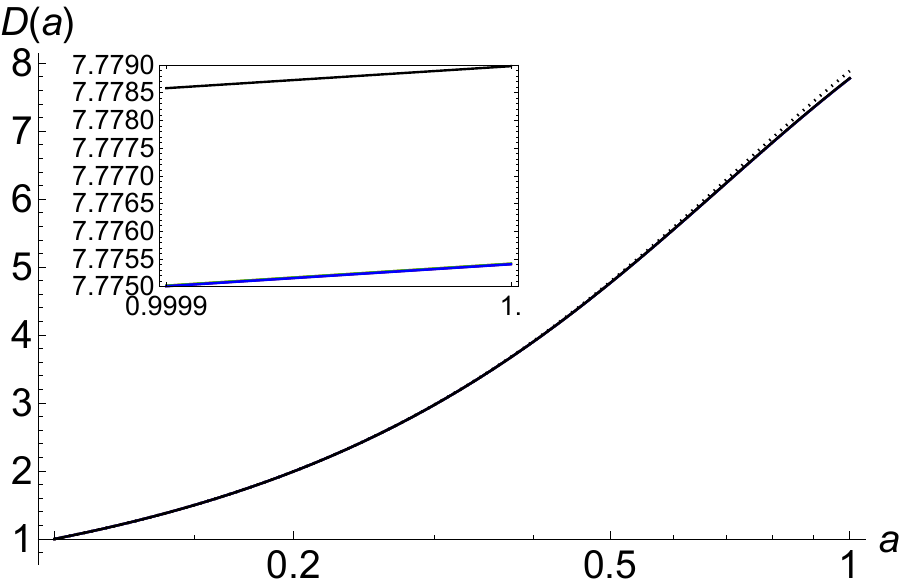}
    \label{fig:linder_CC_SN_BAO_TRGB}}
    \hfill
    \hspace{0.06\textwidth}
    \subfloat
    {\includegraphics[height=0.18\textwidth]{Plots/general_legend.pdf}}
    \hspace{0.06\textwidth}
    \caption{Growth factor plots for the Linder model. Red line corresponds to the growth factor solution at radiation-matter equality, the green represents the solution at the point just before subhorizon limit applies ($k_{\text{lim}}$), and the blue line is the subhorizon solution. Black, solid line is the growth factor at $\Lambda$CDM limit obtained for the model as $\beta_{2} \rightarrow +\infty$ ($\Lambda\text{CDM}_{\text{Model}}$), and the dotted line is the $\Lambda$CDM solution for the results obtained in the Planck collaboration ($\Lambda\text{CDM}_{\text{Planck}}$)~\cite{Planck:2018vyg}. The yellow curve, where $k = 2 k_{\text{lim}}$, is not visible in the scales presented as these values are well within the subhorizon limit and can be depicted by the blue curve.}
    \label{fig:linder}
\end{figure}

The results for the Linder model are tabulated in Table~\ref{tab:linder}. Across the data sets included, the growth index is within $0.5601 \leq \gamma \leq 0.5659$, where CC+SN+BAO model, in particular, contributes to the upper end of this range when considering up to 4~d.p. When looking into the $k$ values for which the subhorizon limit is applicable, the majority of them is below the radiation-matter equality value $k_{\text{eq}}$, apart from, once again, CC+SN+BAO. These results correspond to the growth factor behaviour where, for  a large part of the cases, the growth factor at $k_{\text{eq}}$ (red line) is not visible in the plot as these values are well-within the subhorizon regime given by Eqs.~(\ref{eq:coeff_Ddeltam_sub}-\ref{eq:coeff_deltam_sub}), while the green curve is visible as it is just outside the subhorizon limit. On the other hand, in Fig.~\ref{fig:linder_CC_SN_BAO}, all values $k \leq 119.26 H_{0}$ yield to a slight deviation from the subhorizon limit. Above values, represented by a yellow line, can be encapsulated within the subhorizon plot (blue line). Additionally, the model results are lower than the respective $\Lambda$CDM limit.

For the rest of the models, the subhorizon limit can be considered appropriate for any $k > k_{\text{eq}}$, since the growth index listed in Table~\ref{tab:linder} can be attained for such values. For this reason, see the corresponding plots in Fig.~\ref{fig:linder}, the growth factor solution for the radiation-matter equality is not visible, as it can be described by the subhorizon limit where $\frac{G_{\text{eff}}}{G} = \frac{1}{1+f_{T}}$, as obtained in Eqs.~(\ref{eq:coeff_Ddeltam_sub}-\ref{eq:coeff_deltam_sub}). Due to the larger values of $\beta_{2}$, its exponential behaviour  yields to solution closer to the $\Lambda$CDM limit. This can be seen by the absence of deviation of the growth factor on a larger scale, in contrast with that of the CC+SN+BAO model in Fig.~\ref{fig:linder_CC_SN_BAO}. Additionally, in the data set of CC+SN+BAO+TRGB in Fig.~\ref{fig:linder_CC_SN_BAO_TRGB}, $k_{\text{subhorizon}} \approx k_{\text{eq}}$, hence making the green curve difficult to differentiate from the subhorizon limit (blue) plot without consider very small orders of magnitude.

\begin{table}[h!]
\footnotesize
    \centering
    \begin{tabular}{|c|>{\centering\arraybackslash}p{1cm}|>{\centering\arraybackslash}p{1cm}|>{\centering\arraybackslash}p{1cm}|>{\centering\arraybackslash}p{1.8cm}|>{\centering\arraybackslash}p{1.8cm}|>{\centering\arraybackslash}p{1.8cm}|}
    \hline
    \multicolumn{7}{|c|}{Linder Model}\\ \hline
        Data Set & $\frac{1}{\beta_{2}}$ & $H_{0}$ & $\Omega_{m0}$ & $k_{\text{eq}} / H_{0}$ & $k/H_{0}$ & $\gamma$
        \\ \hline\hline
        CC + SN & $0.101$ & $68.7$ & $0.298$ &  $45.9824$ & $> 15.79$ & $0.5603$ 
        \\
        CC + SN + R19 & $0.088$ & $71.4$ & $0.283$  & $43.6679$ & $>15.05$ & $0.5609$
        \\
        CC + SN + HW & $0.096$ & $71.0$ & $0.285$ & $43.9764$ & $>16.82$ & $0.5608$ 
        \\
       CC + SN + TRGB & $0.088$ & $69.2$ & $0.296$  & $45.6738$ & $>24.25$ & $0.5603$ 
        \\
        CC + SN + BAO & $0.220$ & $66.9$ & $0.294$  & $45.3652$  & $>119.26$ & $0.5659$
        \\
        CC + SN + BAO + R19 & $0.079$ & $68.71$ & $0.300$ & $46.2910$ &  $>6.64$ & $0.5601$
        \\
        CC + SN + BAO + HW & $0.076$ & $68.58$ & $0.300$ & $46.2910$ & $>5.61$ & $0.5601$
        \\
       CC + SN + BAO + TRGB & $0.128$ & $67.7$ & $0.297$  & $45.8281$ & $>45.00$ & $0.5606$
        \\\hline
    \end{tabular}
    \caption{Linder model results for different data sets constraints. The second column corresponds to the $\beta_{2}$ parameter in the models, third column corresponds to the $H_{0}$ value and fourth column gives the $\Omega_{m0}$ value. For each set, the $k$-cutoff value at matter-radiation equality is determined as $k_{\text{eq}}$ in the fourth column. The fifth column gives the minimum value for $k$ in order for the M\'{e}sz\'{a}ros equation to start behaving like the subhorizon limit, particularly determined by the constant $\gamma$ value in the sixth column up to 4~d.p.}
    \label{tab:linder}
\end{table}

\subsection{\texorpdfstring{$f_3(T)$}{}: Exponential Model}\label{subsec:exponential}

As a variant of the Linder model, the Exponential model in $f(T)$ does not counter the square root that appears in the torsion scalar expression of the Hubble parameter, and so provides a distinct cosmic evolution albeit with keeping important features of the exponential form of the Linder model. Moreover, this is analogous to the exponential model of $f(\lc{R})$ gravity~\cite{Linder:2010py}, and is defined as
\begin{align}
    f_{3}(T) = -\alpha_{3} T_{0} \left(1 - \exp{\left(- \beta_{3} \frac{T}{T_{0}}\right)}\right)\,,
\end{align}
where once again $T_{0} = -6{H_{0}}^{2}$, $\beta_{3}$ and $\alpha_{3}$ are constants, with the latter determined through the Friedmann equation~\eqref{eq:background_friedmann_1} at current time 
\begin{align}
    \alpha_{3} = \frac{e^{\beta_{3}} (1-\Omega_{m0})}{e^{\beta_{3}}-1 - 2\beta_{3} }\,,
\end{align}
leading to
\begin{align}
    h(a)^{2} = \Omega_{m0} a^{-3} + (1-\Omega_{m0}) \frac{\left( 
1 - e^{-\beta_{3}h(a)^{2}} (1 + 2 \beta_{3} h(a)^{2}) \right) }{1 - e^{-\beta_{3}} (1 + 2\beta_{3})}\,.
\end{align}
Similar to Linder model case~(\ref{subsec:f2_model}), as $\beta_{3} \rightarrow +\infty$ the model approaches the $\Lambda$CDM limit. 

\begin{figure}[h!]
    \centering
    \subfloat[Subhorizon Limit]
    {\includegraphics[width=0.3\textwidth]{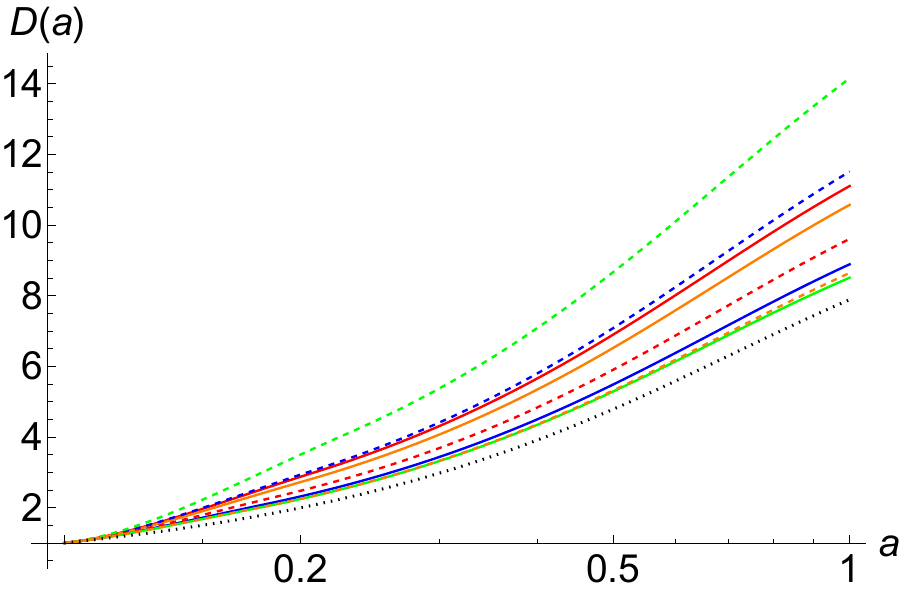}
    \label{fig:exp_subhorizon}}
    \hfill
    \subfloat[$\Lambda$CDM Limit]
    {\includegraphics[width=0.3\textwidth]{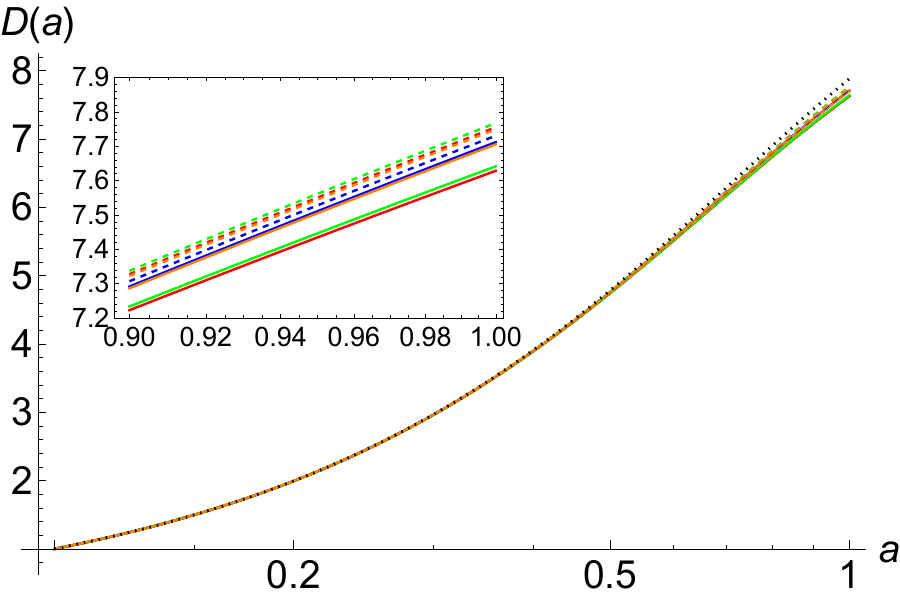}
    \label{fig:exp_lambdaCDM}}
    \hfill
    \subfloat
    {\includegraphics[width=0.2\textwidth]{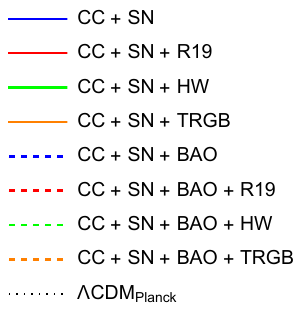}}
    \caption{Growth factor plots for the Exponential model. Fig.~\ref{fig:exp_subhorizon} are the different data set models evaluated at subhorizon limit and Fig.~\ref{fig:exp_lambdaCDM} are the corresponding $\Lambda$CDM limits along the the $\Lambda$CDM result with the values obtained in the Planck collaboration ($\Lambda\text{CDM}_{\text{Planck}}$)~\cite{Planck:2018vyg}.}
    \label{fig:exponential}
\end{figure}

For the list of results obtained from data set combinations, the limits obtained in Eqs~(\ref{eq:coeff_Ddeltam_sub}-\ref{eq:coeff_deltam_sub}) for subhorizon can be considered appropriate for any value $k \geq k_{\text{eq}}$, clearly given by the low values of $k$ in the sixth column of Table~\ref{tab:exponential} in order to obtain $\gamma$ value equivalent up to 4~d.p. For this reason, $k$-dependencies can only be spotted when considering very low orders of magnitudes. Additionally, growth index values are within the range $0.5603 \leq \gamma \leq 0.5613$. In Figs~\ref{fig:exponential}, we compare the subhorizon limit of different data sets presented in Table~\ref{tab:exponential}. The distinction in the growth factor for different data set combinations is amplified with the inclusion of the differing $\beta_{3}$ constant parameter. This can be seen in the difference between the subhorizon plot in Fig.~\ref{fig:exp_subhorizon} and $\Lambda$CDM plots in Fig.~\ref{fig:exp_lambdaCDM}. In particular, the model CN+SN+BAO+HW deviates the most from the $\Lambda$CDM plot, in contrast with it respective model without BAO results.

\begin{table}[h!]
\footnotesize
    \centering
    \begin{tabular}{|c|>{\centering\arraybackslash}p{1cm}|>{\centering\arraybackslash}p{1cm}|>{\centering\arraybackslash}p{1cm}|>{\centering\arraybackslash}p{1.8cm}|>{\centering\arraybackslash}p{1.8cm}|>{\centering\arraybackslash}p{1.8cm}|}
    \hline
    \multicolumn{7}{|c|}{Exponential Model}\\ \hline
    Data Set & $\frac{1}{\beta_{3}}$ & $H_{0}$ & $\Omega_{m0}$ & $k_{\text{eq}} / H_{0}$ & $k/H_{0}$ & $\gamma$
        \\ \hline\hline
        CC + SN & $0.067$ & $69.6$ & $0.286$ & $44.1308$ & $>0.66$ & $0.5607$
       \\
        CC + SN + R19 & $0.042$ & $72.0$ & $0.273$ &  $42.1248$ & $>0.06$ & $0.5613$
        \\
       CC + SN + HW & $0.070$ & $71.5$ & $0.275$  & $42.4334$ & $>1.31$ & $0.5613$
        \\
        CC + SN + TRGB & $0.048$ & $69.7$ & $0.285$ & $43.9764$ & $>0.30$ & $0.5608$
        \\
        CC + SN + BAO & $0.043$ & $67.35$ & $0.289$ &  $44.5937$  & $>0.35$ & $0.5606$
        \\
       CC + SN + BAO + R19 & $0.059$ & $68.7$ & $0.293$ &  $45.2109$ & $>1.19$ & $0.5604$  
        \\
       CC + SN + BAO + HW & $0.034$ & $68.52$ & $0.295$ & $45.5195$ & $>0.01$ & $0.5603$
      \\
        CC + SN + BAO + TRGB & $0.074$ & $67.79$ & $0.292$ &  $45.0566$ & $>2.34$ & $0.5605$
       \\\hline
    \end{tabular}
    \caption{Exponential model results for different data sets constraints. The second column corresponds to the $\beta_{3}$ parameter in the models, the third column corresponds to the $H_{0}$ value and the fourth column gives the $\Omega_{m0}$ value. For each set, the $k$-cutoff value at matter-radiation equality is determined as $k_{\text{eq}}$ in the fourth column. The fifth column gives the minimum value for $k$ in order for the M\'{e}sz\'{a}ros equation to start behaving like the subhorizon limit, particularly determined by the constant $\gamma$ value in the sixth column up to 4~d.p.}
    \label{tab:exponential}
\end{table}

\subsection{\texorpdfstring{$f_4(T)$}{}: Logarithmic Model}\label{subsec:log}

The Logarithmic model~\cite{Bamba:2012cp} has some similar features as self-accelerating cosmologies in DGP braneworld models \cite{Dvali:2000hr}, and is given by
\begin{align}\label{eq:log_function}
    f_{4}(T) = -\alpha_{4} T_{0} \sqrt{\frac{T}{\beta_{4} T_{0}}} \ln\left(\frac{\beta_{4} T_{0}}{T}\right)\,,
\end{align}
where $\alpha_{4}$ and $\beta_{4}$ are the model parameters, the former given by
\begin{align}
    \alpha_{4} = \frac{\sqrt{\beta_{4}}(1 - \Omega_{m0})}{2} \,,
\end{align}
such that
\begin{align} \label{eq:log_friedmann}
    h(a)^{2} = \Omega_{m0} a^{-3} + (1-\Omega_{m0}) h(a)\,.
\end{align}
It should be noted, that the model does not retrieve $\Lambda$CDM cosmology but it does correspond to DGP at a background level~\cite{Dvali:2000hr,Deffayet:2000uy} while it is not necessarily the case at a perturbative level. 

In the logarithmic case, we will take a different approach as no constraint was obtained for the $\beta_{4}$ parameter in Ref.~\cite{Briffa:2021nxg} as it does not play a role in the Friedman background equation~\eqref{eq:log_friedmann} while investigation of growth factor and index still rely on this value. %
We can determine some constraints of the $\beta_{4}$ value from the model, and definitions of the growth factor and the growth index. From Eq.~\eqref{eq:log_function}, the logarithm should be positive definite. Given the positive nature of $\tfrac{T_0}{T}$, we must impose that $\beta_{4} > 0$. Next, consider the  M\'{e}sz\'{a}ros equation~\eqref{eq:meszaros_eq_in_t}, where both its general~(\ref{eq:coeff_Ddeltam}-\ref{eq:coeff_deltam}) and subhorizon limit~(\ref{eq:coeff_Ddeltam_sub}-\ref{eq:coeff_deltam_sub}) forms are undefined at $1 + f_{T} = 0$ . Hence, an undetermined result is obtained, thus it holds for when 
\begin{align}\label{eq:beta4_Geff}
    \beta_{4} \neq h(a)^{2} \exp{\left( 2 - \frac{4 h(a)}{1-\Omega_{m0}}\right)}\,,
\end{align}
where the condition $1-\Omega_{m0} \neq 0$ is consistent with the data set values listed in Table~\ref{tab:log0}. The dependency on $a$ shows that this $\beta_{4}$ value is undefined for a range of values as the Universe evolves during $0.1 \leq a \leq 1$. The solution of $h(a)$ for $0.1 \leq a \leq 1$ and the value $\Omega_{m0}$ depend on the data set considered. Thus, we can apply the constraints for $H_{0}$ and $\Omega_{m0}$ listed in Table~\ref{tab:log0}. It should be noted that the range of $\beta_{4}$ values which are undefined when Eq.~\eqref{eq:beta4_Geff} does not hold because such values correspond to negative constant growth index $\gamma$ values in Eq.~\eqref{eq:growth_index_differential}.
\begin{table}[h!]
\footnotesize
    \centering
    \begin{tabular}{|c|>{\centering\arraybackslash}p{1cm}|>{\centering\arraybackslash}p{1cm}|>{\centering\arraybackslash}p{7cm}|}
    \hline
        \multicolumn{4}{|c|}{Logarithmic Model} \\ \hline
        Data Set & $H_{0}$ & $\Omega_{m0}$ & $\beta_{4}^{a = 0.1} \leq \beta_{4} \leq \beta_{4}^{a = 1}$ 
        \\ \hline\hline
        CC + SN & $68.8$ & $0.207$ & $6.45814 \times 10^{-30} \leq \beta_{4} \leq 0.0476374$ 
        \\
        CC + SN + R19 & $71.4$ & $0.194$ & $1.90723 \times 10^{-28} \leq \beta_{4} \leq 0.0516751$ 
        \\
        CC + SN + HW & $71.0$ & $0.195$& $1.47207 \times 10^{-28} \leq \beta_{4} \leq 0.0513575$ 
        \\
       CC + SN + TRGB & $69.2$ & $0.205$ & $1.09022 \times 10^{-29}\leq \beta_{4} \leq 0.0482458$ 
        \\
         CC + SN + BAO  & $60.89$ & $0.252$  & $3.50098 \times 10^{-35} \leq \beta_{4} \leq 0.0351689$ 
        \\
        CC + SN + BAO + R19 & $62.77$ & $0.261$ & $2.81590 \times 10^{-36} \leq \beta_{4} \leq 0.0329515$ 
        \\
        CC + SN + BAO + HW & $62.77$ & $0.260$ & $3.37274 \times 10^{-36} \leq \beta_{4} \leq 0.0331934$ 
        \\
       CC + SN + BAO + TRGB & $62.02$ & $0.257$ & $8.67071 \times 10^{-36} \leq \beta_{4} \leq 0.0339258$  \\\hline
    \end{tabular}
    \caption{Range of undefined region of $\beta_{4}$ values determined when Eq.~\eqref{eq:beta4_Geff} does not hold in the M\'{e}sz\'{a}ros equation~\eqref{eq:meszaros_eq_in_t} in the general form~(\ref{eq:coeff_Ddeltam}-\ref{eq:coeff_deltam}) and in the subhorizon limit~(\ref{eq:coeff_Ddeltam_sub}-\ref{eq:coeff_deltam_sub}).}
    \label{tab:log0}
\end{table}%

In addition,  small deviations from the $\Lambda$CDM model are expected if the constraint  $f_{T}\ll 1$ is taken into account. In particular, it is worth saying that this situation works if the coupling condition $G_{eff}\rightarrow G$ holds.
Hence, the constraint presented in Eq.~\eqref{eq:beta4_Geff} is further restricted by requiring
\begin{align} \label{eq:beta4_fT}
    \beta_{4} \ll h(a)^{2} \exp{\left(2 + \frac{4 h(a)}{1-\Omega_{m0}}\right)}\,.
\end{align}
The results, deep in the matter era (e.g. $a=0.1$), correspond to large order magnitude $\beta_{4}$ values, which exponentially decrease as current time is approached ($a = 1$). In order to have the same $\beta_{4}$ value adequate as evolution of the Universe is underway, we list the minimum value obtained for current times for which $\beta_{4}$ is much smaller than in Table~\ref{tab:log_fT}. Ignoring the much less than inequality symbol, we obtain the $\gamma$ value using Eq.~\eqref{eq:growth_index_differential}. Thus, the requirement for the results obtained through Eq.~\eqref{eq:beta4_fT}, requires the $\gamma$ value to be smaller than that listed in Table~\ref{tab:log_fT}.

\begin{table}[h!]
\footnotesize
    \centering
    \begin{tabular}{|c|>{\centering\arraybackslash}p{1cm}|>{\centering\arraybackslash}p{1cm}|>{\centering\arraybackslash}p{3cm}|>{\centering\arraybackslash}p{3cm}|}
    \hline
        \multicolumn{5}{|c|}{Logarithmic Model} \\ \hline
        Data Set & $H_{0}$ & $\Omega_{m0}$ & $\beta_{4}$ & $\gamma$
        \\ \hline\hline
        CC + SN & $68.8$ & $0.207$ & $\ll 1146.12$ & $< 0.6908$
        \\
        CC + SN + R19 & $71.4$ & $0.194$ & $\ll 1056.57$ & $<0.6860$
        \\
        CC + SN + HW & $71.0$ & $0.195$ & $\ll 1063.10$ & $<0.6863$
        \\
       CC + SN + TRGB & $69.2$ & $0.205$ & $\ll 1131.67$
        & $<0.6900$\\
         CC + SN + BAO  & $60.89$ & $0.252$  & $\ll 1552.46$ & $< 0.7089$
        \\
        CC + SN + BAO + R19 & $62.77$ & $0.261$ & $\ll 1656.93$ & $<0.7127$
        \\
        CC + SN + BAO + HW & $62.77$ & $0.260$ & $\ll 1644.85$ & $<0.7123$
        \\
       CC + SN + BAO + TRGB & $62.02$ & $0.257$ & $\ll 1609.34$ & $<0.7110$  \\\hline
    \end{tabular}
    \caption{Results from the constraint $f_{T} \ll 1$, corresponding to Eq.~\eqref{eq:beta4_fT} are listed in the fourth column. The fifth column contains the $\gamma$ value obtained for the exact value listed in the fourth column. Due to the inequality for the $\beta_{4}$ values, it is observed significant smaller values give a smaller $\gamma$ value, hence for the introduction of the inequality in the fifth column results.}
    \label{tab:log_fT}
\end{table}

Next, we will consider the growth index equation~\eqref{eq:growth_index_differential} in its subhorizon form, which in the Logarthmic model is given by
\begin{align}
    0 =  \Omega_{m0}^{2\gamma} - \frac{6 \Omega_{m0}}{2(1+\Omega_{m0}) + (1-\Omega_{m0}) \ln \beta_{4}} + \Omega_{m0}^{\gamma} \left[2 + h'(1) - \gamma(3 + 2h'(1))\right]\,,
\end{align}
undefined when its denominator $2(1+\Omega_{m0}) + (1-\Omega_{m0})\ln\beta_{4}$ = 0. As Eq.~\eqref{eq:growth_index_differential} is defined at $a(t_0) = 1$, following Eq.~\eqref{eq:beta4_Geff}, the $\beta_4$ at which the growth index differential equation is undefined is given by the upper bound of the fourth column in Table~\ref{tab:log0}. In general, we can solve for $\beta_{4}$ using Eq.~\eqref{eq:growth_index_differential} within the subhorizon limit:
\begin{align} \label{eq:beta4_growthindex}
    \beta_{4} = \exp{\left[ 
    \frac{-6 \Omega_{m0} + 2 \Omega_{m0}^{2\gamma} (1+\Omega_{m0}) - 2 \Omega_{m0}^{\gamma}(1+\Omega_{m0}) \left[-2 + 3 \gamma + (-1+2\gamma)h'(1)\right] }
    { \Omega_{m0}^{\gamma} (-1+\Omega_{m0}) \left[2 + \Omega_{m0}^{\gamma} + h'(1) - \gamma (3 + 2 h'(1))\right] }
    \right]}\,,
\end{align}
where the boundary condition $h(1)=1$ is implemented as this holds for all data set models. Once again, we obtain another undefined region for which the term in the exponent approaches $+\infty$. This is due to the dependency on $\beta_{4}$ for the $\alpha_{4}$ parameter by virtue of the background equations. In particular, it is undefined at
\begin{align}\label{eq:beta4_gamma_undefined}
    2 + \Omega_{m0}^\gamma + h'(1) - \gamma(3+2h'(1)) = 0\,.
\end{align}
Since $h'(1)$ value varies for different data sets, then a constant $\gamma$ has to be obtained for each model. By solving Eq.~\eqref{eq:beta4_gamma_undefined} for $\gamma$, we can obtain a value at which the Logarithmic model does not hold. These results have been summarized in the fourth column of Table~\ref{tab:log1}. Next, we will consider $\gamma$ value to lie within $0 < \gamma < 1$. The behaviour is consistent with studies in $f(\lc{R})$~\cite{Gannouji:2009zz,Tsujikawa:2009ku}, running vacuum models~\cite{Basilakos:2015vra}, DGP gravity~\cite{Wei:2008ig,Fu:2009nr,Gong:2008fh,Linder:2007hg}, Finsler-Randers type cosmology~\cite{Basilakos:2013ij}, along with the GR results for $\gamma = \frac{6}{11}$. The values of $\beta_{4}$ at $\gamma = 0$ are listed in Table~\ref{tab:log1}. As $\gamma$ increases, $\beta_4$ increases to very large orders of magnitude up until it becomes undefined at some growth index, $\gamma^{\text{undefined}}$, where $\gamma^{\text{undefined}}$ is the value obtained from solving Eq.~\eqref{eq:beta4_gamma_undefined}. After this value, $\beta_{4}$ constant drops to zero and slowly increases to $\beta_{4}$ values at $\gamma = 1$. While, values in the range $\beta_{4}^{\gamma = 0} \leq \beta_{4} \leq \beta_{4}^{\gamma = 1}$ satisfy the condition given by Eq.~\eqref{eq:beta4_fT}, the value $\gamma = 1$ is undefined for growth index values given by Table~\ref{tab:log_fT}.
\begin{table}[h!]
\footnotesize
    \centering
    \begin{tabular}{|c|>{\centering\arraybackslash}p{1cm}|>{\centering\arraybackslash}p{1cm}|>{\centering\arraybackslash}p{2cm}|>{\centering\arraybackslash}p{2.5cm}|>{\centering\arraybackslash}p{2.5cm}|}
    \hline
        \multicolumn{6}{|c|}{Logarithmic Model}\\ \hline
        Data Set & $H_{0}$ & $\Omega_{m0}$ & $\gamma^{\text{undefined}}$ & $\beta_{4}^{\gamma = 0}$ & $\beta_{4}^{\gamma = 1}$ 
        \\ \hline\hline
        CC + SN & $68.8$ & $0.207$ & $0.880459$ & $0.0894568$ & $7.57215 \times 10^{-14}$
        \\
        CC + SN + R19 & $71.4$ & $0.194$ & $0.866199$ & $0.0918131$ & $3.67027 \times 10^{-12}$
        \\
        CC + SN + HW & $71.0$ & $0.195$ &  $0.867282$ & $0.0916297$ & $2.81499 \times 10^{-12} $
        \\
       CC + SN + TRGB & $69.2$ & $0.205$ & $0.878239$ & $0.0898153$& $1.47126 \times 10^{-13}$
        \\
         CC + SN + BAO  & $60.89$ & $0.252$  &  $0.933138$ & $0
         0817573$& $2.4102 \times 10^{-26}$
        \\
        CC + SN + BAO + R19 & $62.77$ & $0.261$ & $0.944366$ & $0.0802988$ & $4.48921 \times 10^{-32}$
        \\
        CC + SN + BAO + HW & $62.77$ & $0.260$ & $0.943106$ & $0.0804596$ & $2.55792 \times 10^{-31}$
        \\
       CC + SN + BAO + TRGB & $62.02$ & $0.257$  &
       $0.939345$ & $0.0809438$ & $2.99871 \times 10^{-29}$\\\hline
    \end{tabular}
    \caption{The constraints of $H_{0}$ and $\Omega_{m0}$ for each data set are listed in the second and third columns, respectively. The fourth column contains the value of $\gamma$ upon solving Eq.~\eqref{eq:beta4_gamma_undefined} corresponding to an undefined value of $\beta_{4}$ as given by Eq.~\eqref{eq:beta4_growthindex}. The fifth column correspond to $\beta_{4}$ value for when $\gamma  = 0$. With increasing $\gamma$ values, $\beta_{4}$ value increases to very large order of magnitudes until it reaches the undefined values of $\gamma^{\text{undefined}}$. Beyond $\gamma^{\text{undefined}}$, $\beta_{4}$ value drops to zero and increases once again to the value given by $\beta_{4}^{\gamma = 1}$ found in the sixth column.}
    \label{tab:log1}
\end{table}%
Having obtained some constraints on $\gamma$ values, we will consider the range of values while steering away from the obtained constraints.
Taking into consideration  the $\Lambda$CDM value from the Planck collaboration in Eq.~\eqref{eq:gamma_lambdaCDM_Planck}, we will obtain a value for $\beta_{4}$ at the end of the range $0.5 \leq \gamma \leq 0.6$. For both ends, the $k$ value at which the subhorizon limit attains the respective $\gamma$ is included in Table~\ref{tab:log}. For the case $\gamma = 0.5$, the $\beta$ parameter is within the range $1.74 \leq \beta_{4} \leq 1.99$, while, for $\gamma = 0.6$, the range is at higher values of $15.51 \leq \beta_{4} \leq 16.31$. The $k$ value at which the subhorizon limit applies is above $k_{\text{eq}}$ for all data sets; in particular, $k > 2.2 k_{\text{eq}}$, where the minimum value is given by the CC+SN+BAO+R19 model for the assumption of $\gamma = 0.5$. For $k > 34.78 k_{\text{eq}}$ and all data sets, irrespective of the growth index value in the range $0.5\leq\gamma\leq0.6$, the model is well within the subhorizon limit.

\begin{table}[h!]
\footnotesize
    \centering
    \begin{tabular}{|c|>{\centering\arraybackslash}p{1cm}|>{\centering\arraybackslash}p{1cm}|>{\centering\arraybackslash}p{1.5cm}|>{\centering\arraybackslash}p{1cm}|>{\centering\arraybackslash}p{1.5cm}|>{\centering\arraybackslash}p{1cm}|>{\centering\arraybackslash}p{1.5cm}|}
    \hline
        \multicolumn{8}{|c|}{Logarithmic Model}\\ \hline
        Data Set & $H_{0}$ & $\Omega_{m0}$ & $k_{\text{eq}} / H_{0}$ &  $\beta_{4}^{\gamma = 0.5}$ & $k^{\gamma=0.5}/H_{0}$ & $\beta_{4}^{\gamma = 0.6}$ & $k^{\gamma=0.6}/H_{0}$
        \\ \hline\hline
        CC + SN & $68.8$ & $0.207$ & $31.9410$ & $1.75$ & $>119.20$ &$16.16$ & $>921.40$
        \\
        CC + SN + R19 & $71.4$ & $0.194$ & $29.9350$  & $1.69$ & $>107.44$ & $16.31$ & $>1041.72$
        \\
        CC + SN + HW & $71.0$ & $0.195$& $30.0893$ & $1.69$ & $>82.62$ & $16.30$ & $>808.05$ 
        \\
       CC + SN + TRGB & $69.2$ & $0.205$ & $31.6324$  & $1.74$ & $>107.14$ & $16.18$ & $>569.75$ 
        \\
         CC + SN + BAO  & $60.89$ & $0.252$  & $38.8846$ & $1.96$ & $>98.93$ & $15.62$ & $>785.53$
        \\
        CC + SN + BAO + R19 & $62.77$ & $0.261$ & $40.2733$ & $1.99$ &  $>88.75$ & $15.51$ & $>628.97$
        \\
        CC + SN + BAO + HW & $62.77$ & $0.260$ & $40.1190$ & $1.99$ & $>266.85$ & $15.53$ & $>645.22$
        \\
       CC + SN + BAO + TRGB & $62.02$ & $0.257$ & $39.6561$ & $1.98$ & $>128.87$ & $15.56$ & $>848.85$
        \\\hline
    \end{tabular}
    \caption{Logarithmic model results for different data sets constraints. The second and third columns correspond to $H_{0}$ and $\Omega_{m0}$ values, respectively. For each set, the $k$-cutoff value at matter-radiation equality is determined as $k_{\text{eq}}$ in the fourth column. For $\beta_{4}$ subhorizon value,  the $\gamma = 0.5$ and $\gamma = 0.6$ values are given by the  fifth and seventh columns, respectively. Their corresponding minimum values for $k$, in order for the M\'{e}sz\'{a}ros equation to start behaving like the subhorizon limit, are given by the sixth and eighth columns.}
    \label{tab:log}
\end{table}

\begin{figure}[h!]
    \centering
    \subfloat[CC + SN]
    {\includegraphics[width=0.3\textwidth]{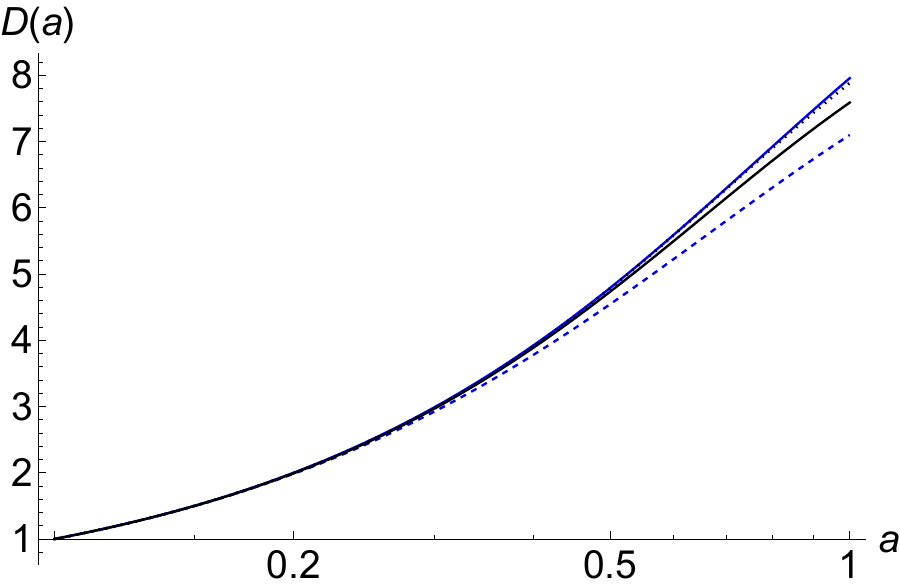}
    \label{fig:log_CC_SN}}
    \hfill
    \subfloat[CC + SN + R19]{\includegraphics[width=0.3\textwidth]{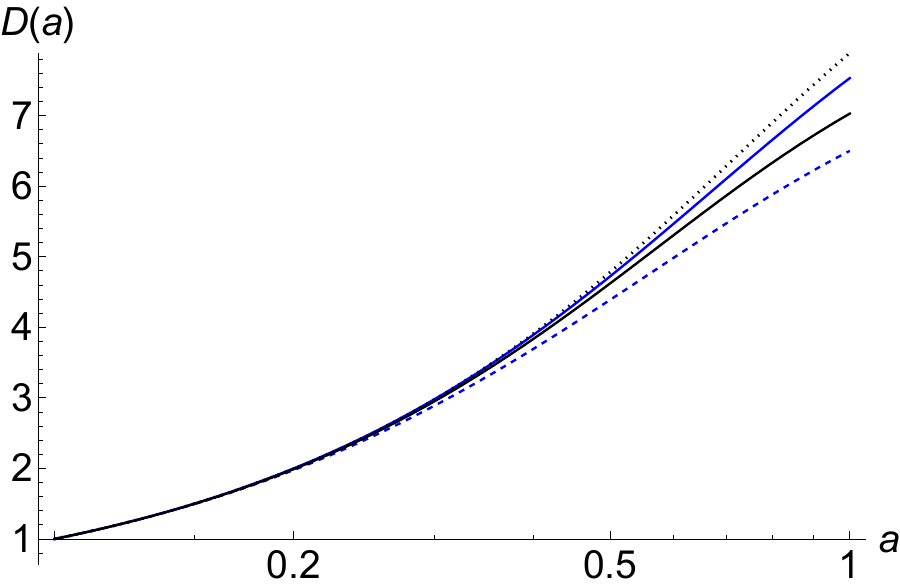}
    \label{fig:log_CC_SN_R19}}
    \hfill
    \subfloat[CC + SN + HW]{\includegraphics[width=0.3\textwidth]{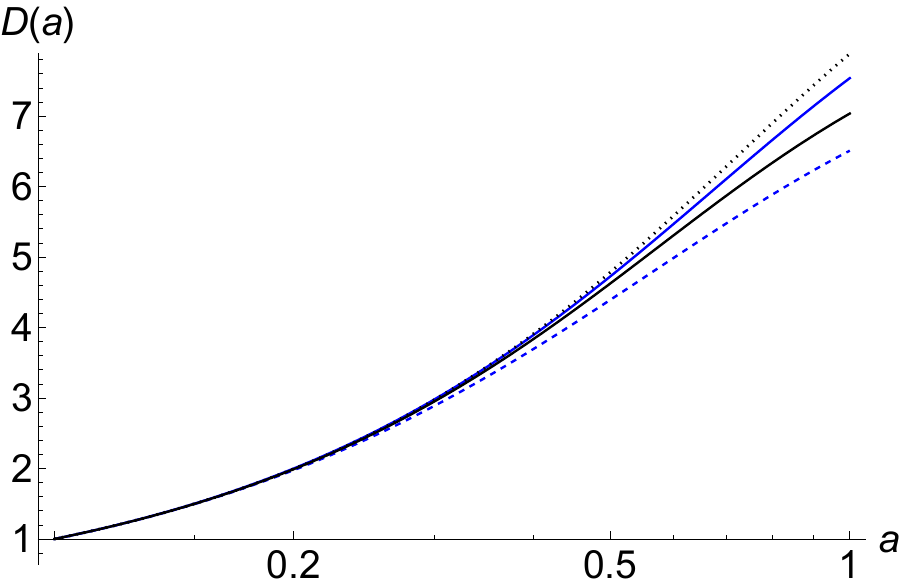}
    \label{fig:log_CC_SN_HW}}
    \vskip 0.1cm
    \subfloat[CC + SN + TRGB]    {\includegraphics[width=0.3\textwidth]{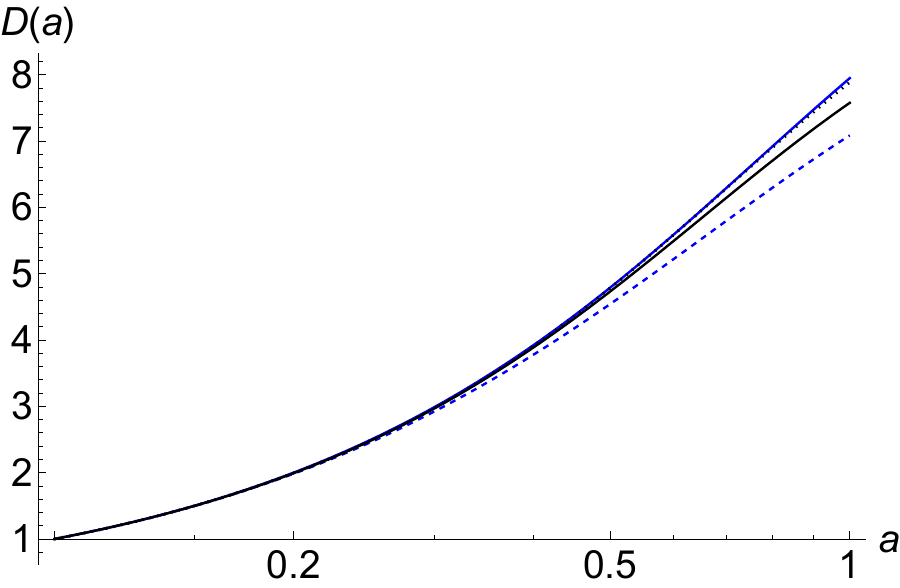}
    \label{fig:log_CC_SN_TRGB}}
    \hfill
    \subfloat[CC + SN + BAO]{\includegraphics[width=0.3\textwidth]{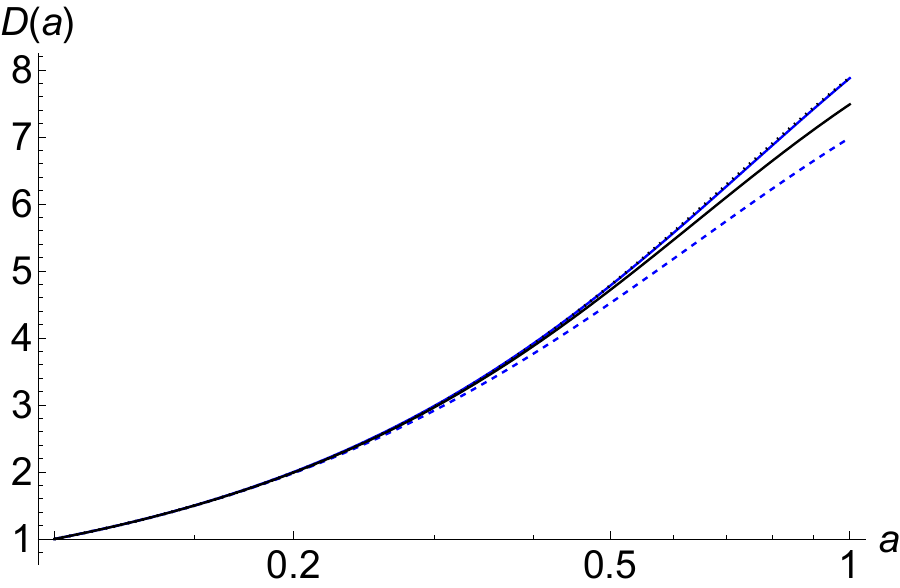}
    \label{fig:log_CC_SN_BAO}}
    \hfill
    \subfloat[CC + SN + BAO + R19]{\includegraphics[width=0.3\textwidth]{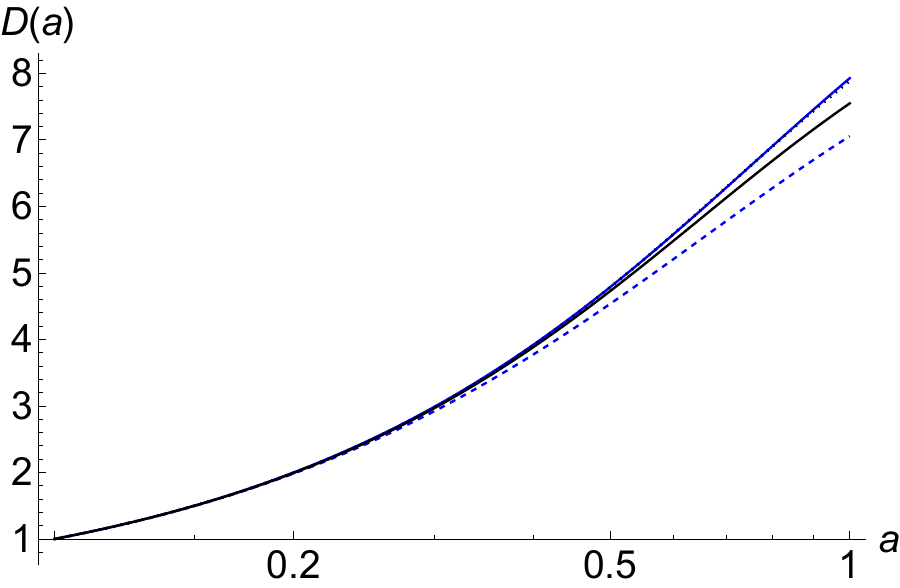}
    \label{fig:log_CC_SN_BAO_R19}}
    \vskip 0.1cm
    \subfloat[CC + SN + BAO + HW]{\includegraphics[width=0.3\textwidth]{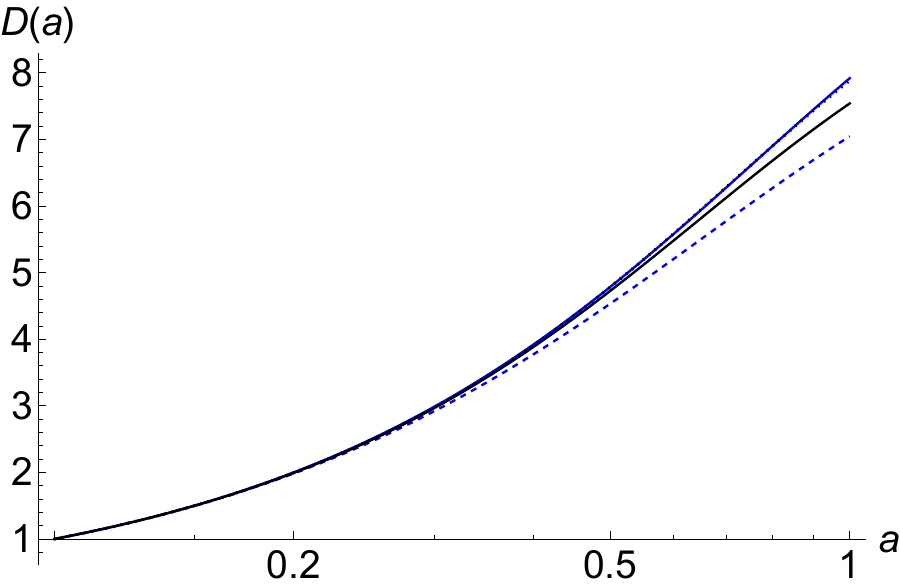}
    \label{fig:log_CC_SN_BAO_HW}}
    \hfill
    \subfloat[CC + SN + BAO + TRGB]{\includegraphics[width=0.3\textwidth]{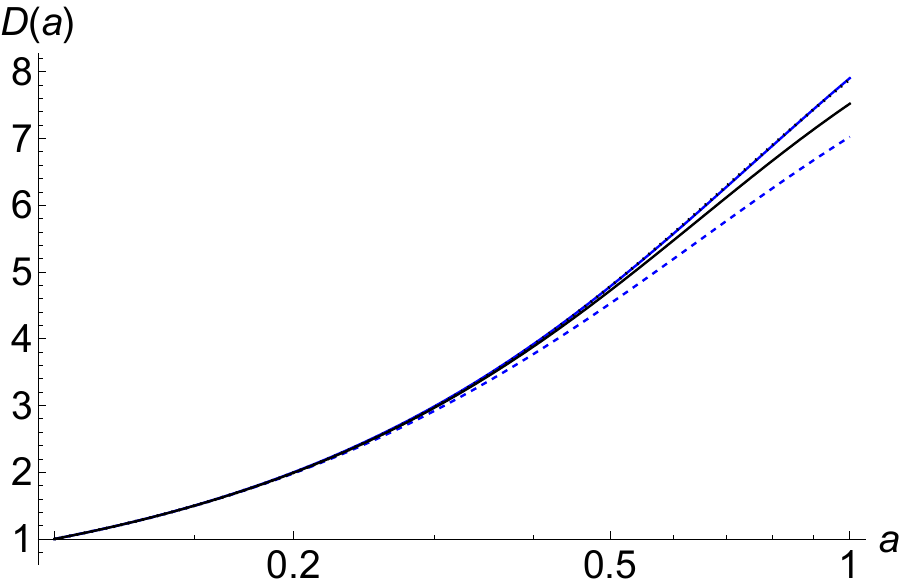}
    \label{fig:log_CC_SN_BAO_TRGB}}
    \hfill
    \hspace{0.07\textwidth}
    \subfloat
    {\includegraphics[width=0.16\textwidth]{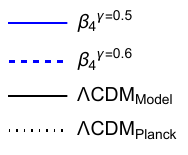}}
    \hspace{0.07\textwidth}
    \caption{Growth factor plots for the Logarithmic model. Blue solid line represent the growth factor solution for $\gamma = 0.5$ and blue dotted line is the result for $\gamma = 0.6$, both at the subhorizon limit. These plots are obtained by retrieving the respective $\beta_{4}$ constant at which the specific $\gamma$ value at the subhorizon limit is obtained. Black solid line is the growth factor for a constant model $f(T) = c$ using parameters from the Logarithm model ($\Lambda\text{CDM}_{\text{Model}}$), and the dotted line is the $\Lambda$CDM solution for the results obtained by the Planck collaboration ($\Lambda\text{CDM}_{\text{Planck}}$)~\cite{Planck:2018vyg}.}
    \label{fig:log}
\end{figure}

 Fig.~\ref{fig:log} represents the growth factor for each data set of the Logarithmic model. For each case, we depict the growth for $\gamma = 0.5$, $\gamma = 0.6$, $\Lambda$CDM using data from the data set model ($\Lambda\text{CDM}_{\text{Model}}$)and $\Lambda$CDM using values from the Planck collaboration ($\Lambda\text{CDM}_{\text{Planck}}$). In all cases, the $\gamma = 0.5$ case is above the $\Lambda\text{CDM}_{\text{Model}}$, while $\gamma = 0.6$ scenario is below it. Figs.~(\ref{fig:log_CC_SN}-\ref{fig:log_CC_SN_TRGB}) correspond to the models without the BAO data sets, which experience a slower growth than the $\Lambda\text{CDM}_{\text{Planck}}$ scenario. On the other hand, the inclusion of BAO data sets gives an overall faster growth in comparison with the previous plots, as seen in Figs~(\ref{fig:log_CC_SN_BAO}-\ref{fig:log_CC_SN_BAO_TRGB}). In particular, BAO models at $\gamma=0.5$ are slightly higher than the $\Lambda\text{CDM}_{\text{Planck}}$, while $\Lambda\text{CDM}_{\text{Model}}$ and the case for $\gamma = 0.6$ remain lower than $\Lambda\text{CDM}_{\text{Planck}}$.


\subsection{\texorpdfstring{$f_5(T)$}{}: Hyperbolic-Tangent Model}\label{subsec:tanh}

Finally, we consider the Hyperbolic-Tangent model~\cite{Wu:2010av}, which has interesting asymptotes that may offer expansion profiles resonating with measurements of the physical Universe. It is given by
\begin{align}
    f_{5}(T) = -\alpha_{5} (-T)^{\beta_{5}} \tanh\left(\frac{T_{0}}{T}\right)\,,
\end{align}
where
\begin{align}
    \alpha_{5} = \frac{(6 H_{0}^{2})^{1-\beta_{5}}(1-\Omega_{0})}{ \tanh(1) (2 \beta_{5} - 1) - 2 \sech^{2}(1)}\,,
\end{align}
and $\beta_{5}$ is a constant, resulting in the Friedman equation
\begin{align}
    h(a)^{2} = \Omega_{m0} a^{-3} + \frac{(1 - \Omega_{m0})h(a)^{2(-1+\beta_{5})} \left( - 2 \sech^{2}(h(a)^{-2}) + (-1+2\beta_{5}) h(a)^{2} \tanh(h(a)^{-2})\right)}{ \tanh(1) (2 \beta_{5} - 1) - 2 \sech^{2}(1)}\,,
\end{align}
for which the model does not attain the $\Lambda$CDM limit. 

As mentioned previously, the $\Lambda$CDM model with $H_{0} = 67.4\,\text{km}\,\text{s}^{-1}\,\text{Mpc}^{-1}$ and $\Omega_{m0} = 0.315$ is used for comparison along with the case of $f(T) = c$, where $c$ is a constant with the parameter constraints of the corresponding data set combinations. Analogous to the power law model~(\ref{subsec:power_law}), $k$ value at which the results coincide with the subhorizon limit for $\gamma$ up to 4~d.p. is always larger than the radiation-matter equality value of $k_{\text{eq}}$. This can be seen from the results in Table~\ref{tab:tanh}, where our subhorizon limit plays a role for when $k > 3.7 k_{\text{eq}}$ at a minimum (CC+SN+BAO) and $k > 8.39 k_{\text{eq}}$ at the higher end (CC+SN+HW). The subhorizon constant growth index is on the lower end, ranging $0.4905 \leq \gamma \leq 0.5193$. The negative $\beta_{5}$ values corresponding models, set without BAO, yield a growth index of $\gamma <0.5$, while the rest of the set with positive $\beta_{5}$ values gives  $\gamma > 0.5$.

\begin{figure}[h!]
    \centering
    \subfloat[CC + SN]{\includegraphics[width=0.3\textwidth]{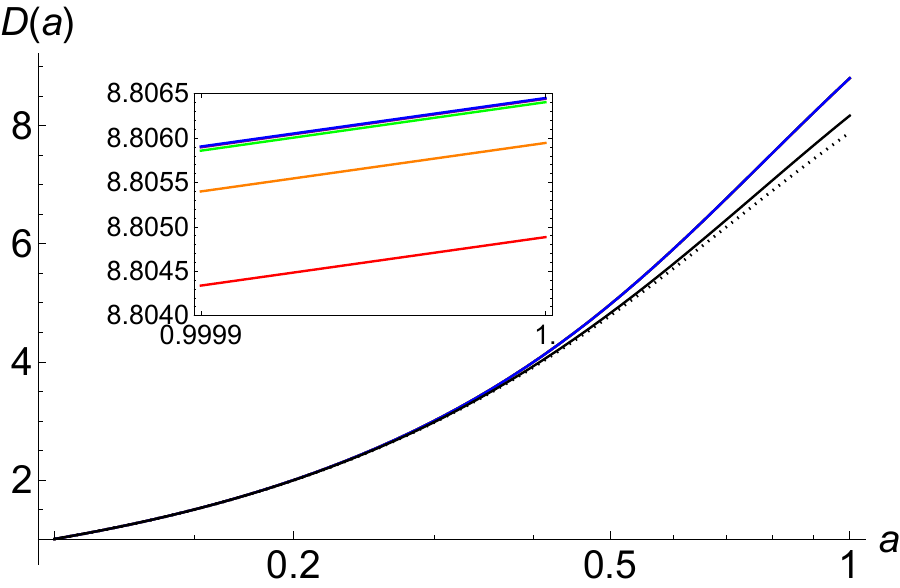}
    \label{fig:tanh_CC_SN}}
    \hfill
    \subfloat[CC + SN + R19]{\includegraphics[width=0.3\textwidth]{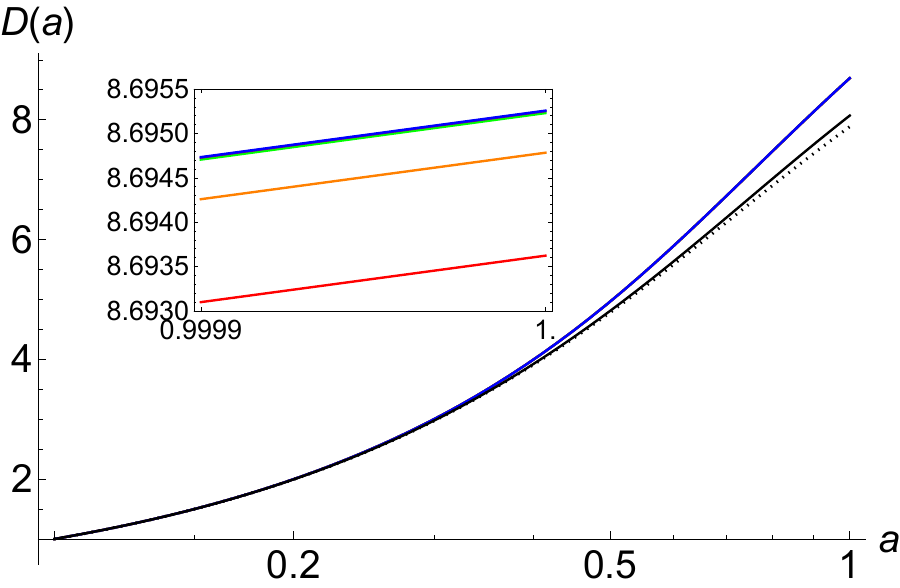}
    \label{fig:tanh_CC_SN_R19}}
    \hfill
    \subfloat[CC + SN + HW]{\includegraphics[width=0.3\textwidth]{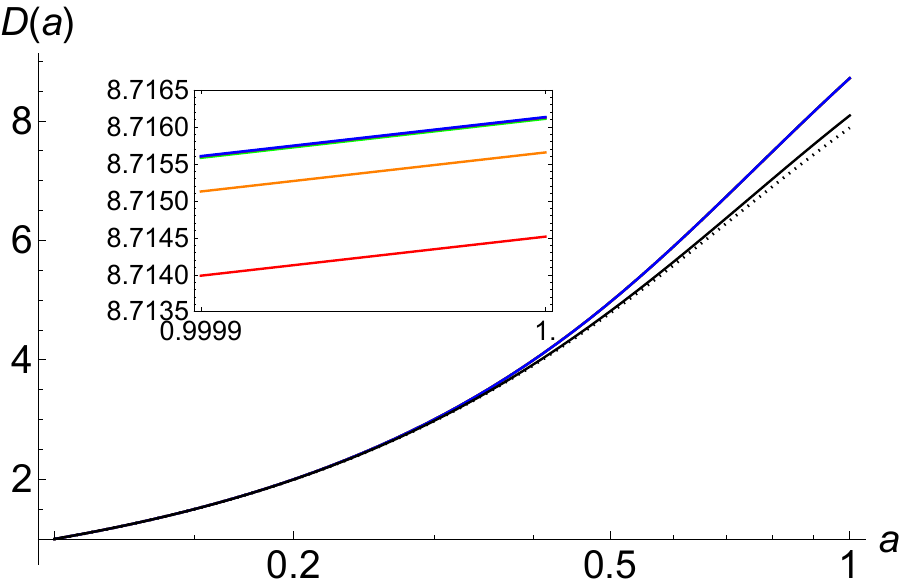}
    \label{fig:tanh_CC_SN_HW}}
    \vskip 0.1cm
    \subfloat[CC + SN + TRGB]{\includegraphics[width=0.3\textwidth]{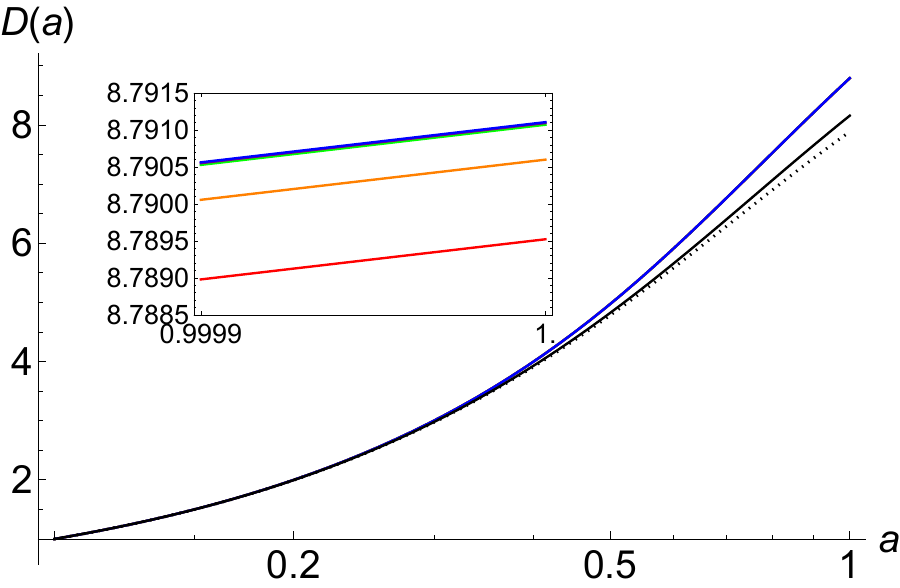}
    \label{fig:tanh_CC_SN_TRGB}}
    \hfill
    \subfloat[CC + SN + BAO]{\includegraphics[width=0.3\textwidth]{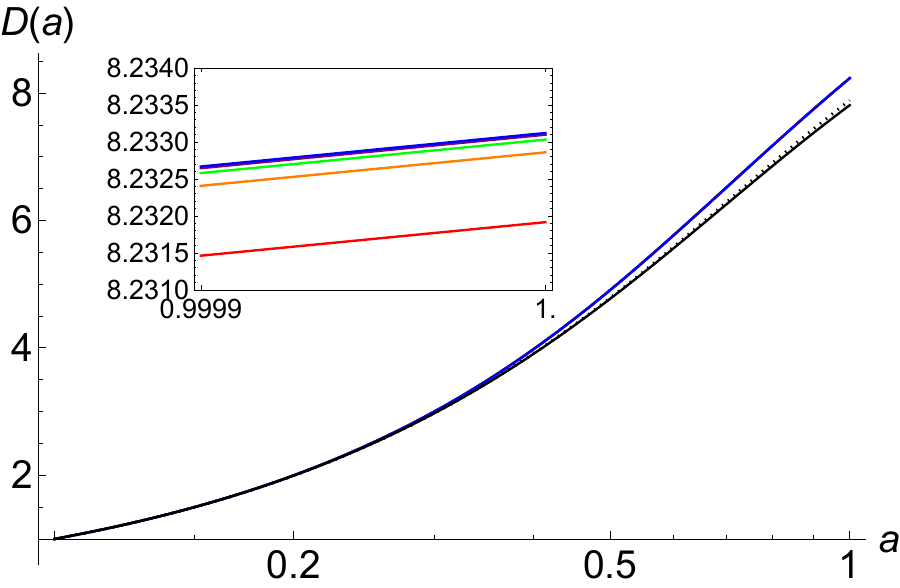}
    \label{fig:tanh_CC_SN_BAO}}
    \hfill
    \subfloat[CC + SN + BAO + R19]{\includegraphics[width=0.3\textwidth]{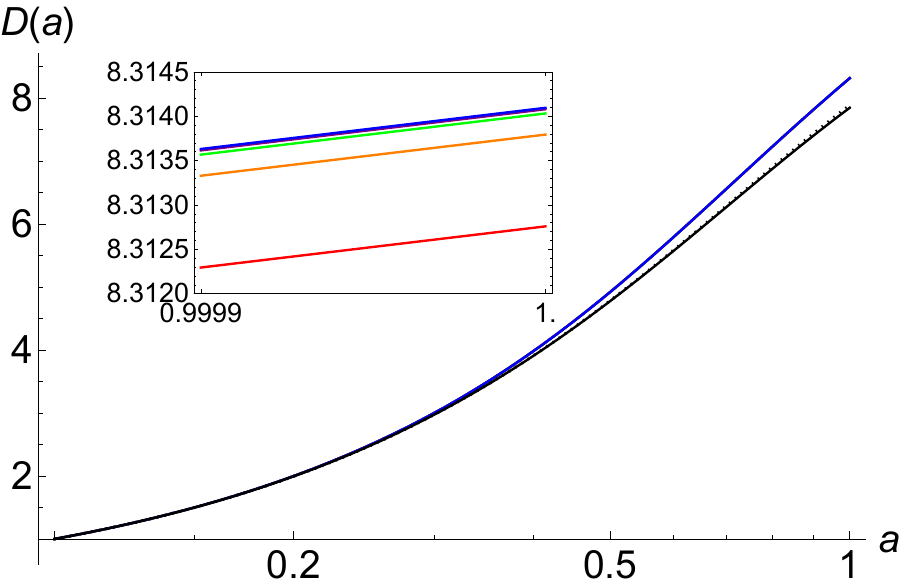}
    \label{fig:tanh_CC_SN_BAO_R19}}
    \vskip 0.1cm
    \subfloat[CC + SN + BAO + HW]{\includegraphics[width=0.3\textwidth]{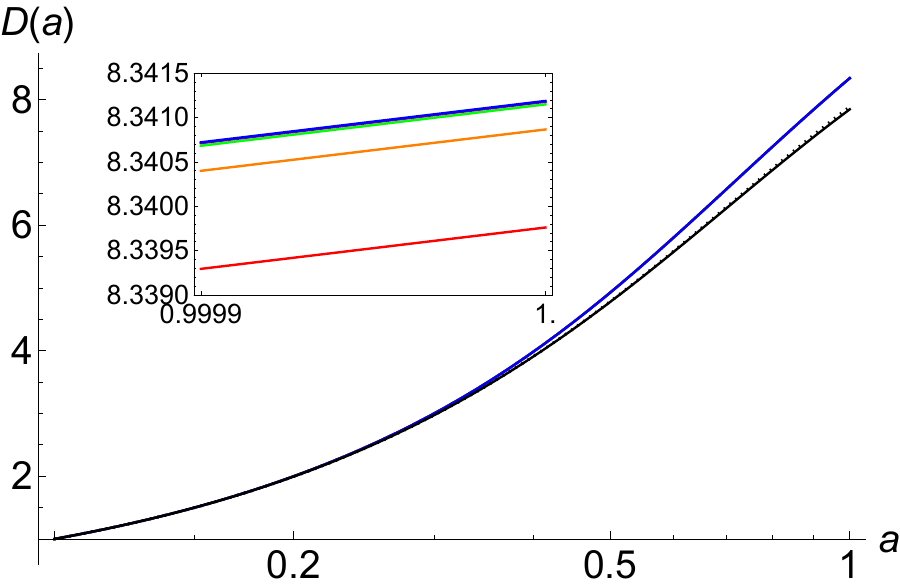}
    \label{fig:tanh_CC_SN_BAO_HW}}
    \hfill
    \subfloat[CC + SN + BAO + TRGB]{\includegraphics[width=0.3\textwidth]{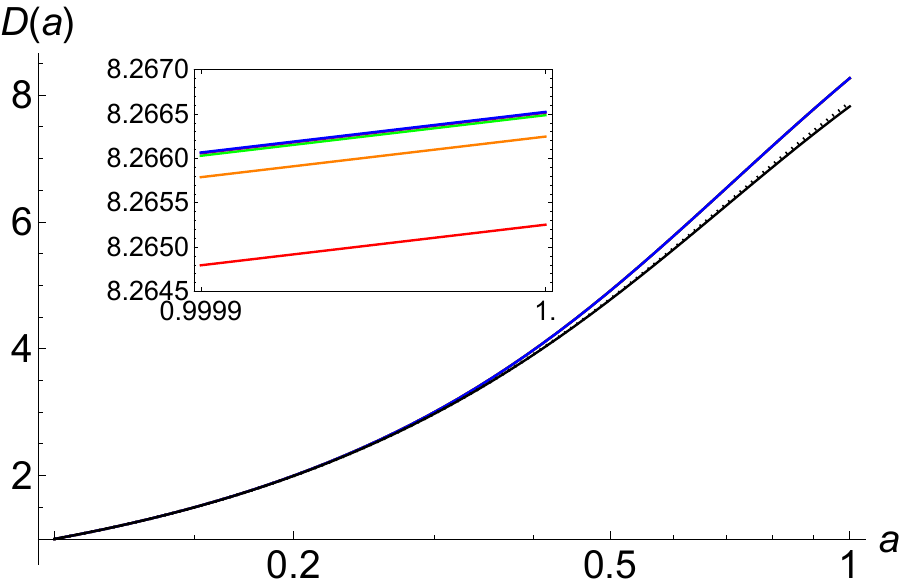}
    \label{fig:tanh_CC_SN_BAO_TRGB}}
    \hfill
    \hspace{0.06\textwidth}
    \subfloat
    {\includegraphics[width=0.18\textwidth]{Plots/general_legend.pdf}}
    \hspace{0.06\textwidth}
    \caption{Growth factor plots for the Hyperbolic Tangent model. The  Red line is the growth factor solution at radiation-matter equality, the green is the solution at the point just before subhorizon limit applies ($k_{\text{lim}}$), and the blue line is the subhorizon solution. The black solid line is the growth factor for $f(T) = c$ using parameters from the hyperbolic tangent model ($\Lambda\text{CDM}_{\text{Model}}$), and the dotted line is the $\Lambda$CDM solution for the results obtained by the Planck collaboration~\cite{Planck:2018vyg} ($\Lambda\text{CDM}_{\text{Planck}}$). The yellow curve, where $k = 2 k_{\text{lim}}$, is not visible in the scales presented as these values are well within the subhorizon limit and can be depicted by the blue curve.}
    \label{fig:tanh}
\end{figure}

The distinction between cases with and without BAO considerations is also highlighted in the growth factor plots in Figs~\ref{fig:tanh}, where in Figs~(\ref{fig:tanh_CC_SN}-\ref{fig:tanh_CC_SN_TRGB}) the difference between the Hyperbolic-Tangent and $\Lambda$CDM growth is larger than those of Figs~(\ref{fig:tanh_CC_SN_BAO}-\ref{fig:tanh_CC_SN_BAO_TRGB}). Additionally, the BAO results, used to plot the $\Lambda$CDM counterpart for a case of $f(T) = c$, are quite close to the $\Lambda$CDM result obtained using values from the Planck collaboration~\cite{Planck:2018vyg}.

\begin{table}[h!]
\footnotesize
    \centering
    \begin{tabular}{|c|>{\centering\arraybackslash}p{1cm}|>{\centering\arraybackslash}p{1cm}|>{\centering\arraybackslash}p{1cm}|>{\centering\arraybackslash}p{1.8cm}|>{\centering\arraybackslash}p{1.8cm}|>{\centering\arraybackslash}p{1.8cm}|}
    \hline
        \multicolumn{7}{|c|}{Hyperbolic Tangent Model}\\ \hline
        Data Set & $\beta_{5}$ & $H_{0}$ & $\Omega_{m0}$ & $k_{\text{eq}} / H_{0}$ & $k/H_{0}$ & $\gamma$
        \\ \hline\hline
        CC + SN & $-0.36$ & $69.2$ & $0.369$ & $56.9379$ & $>327.98$ & $0.4905$
        \\
         CC + SN + R19 & $-0.28$ & $71.8$ & $0.349$ &  $53.8519$ & $>403.54$ & $0.4946$
        \\
       CC + SN + HW & $-0.29$ & $71.3$ & $0.353$  & $54.4691$ & $>456.95$ & $0.4939$
        \\
        CC + SN + TRGB & $-0.35$ & $69.5$ & $0.366$ & $56.4750$ & $>383.85$ & $0.4911$
        \\
        CC + SN + BAO & $0.144$ & $68.4$ & $0.302$ &  $46.5996$  & $>172.47$ & $0.5193$
        \\
       CC + SN + BAO + R19 & $0.079$ & $70.6$ & $0.308$ &  $47.5254$ & $>217.63$ & $0.5142$  
        \\
       CC + SN + BAO + HW & $0.039$ & $70.2$ & $0.308$ & $47.5254$ & $>293.25$ & $0.5118$
      \\
        CC + SN + BAO + TRGB & $0.115$ & $68.9$ & $0.304$ &  $46.9082$ & $>275.94$ & $0.5170$
       \\\hline
    \end{tabular}
    \caption{Hyperbolic-Tangent model results for different data sets constraints. The second column corresponds to the $\beta_{5}$ parameter in the models, the third column corresponds to the $H_{0}$ value, and the  fourth column gives the $\Omega_{m0}$ value. For each set, the $k$-cutoff value at matter-radiation equality is determined as $k_{\text{eq}}$ in the fourth column. The fifth column gives the minimum value for $k$ in order for the M\'{e}sz\'{a}ros equation to start behaving like the subhorizon limit, particularly determined by the constant $\gamma$ value in the sixth column up to 4~d.p.}
    \label{tab:tanh}
\end{table}

\section{Discussion and conclusions} \label{sec:conclu}

The M\'{e}sz\'{a}ros equation~\eqref{eq:meszaros_eq_in_t} provides information regarding CDM perturbations in terms of the fractional matter perturbation defined in Eq.~\eqref{eq:fractional_matter}.It gives insight into the growth profile of large scale structure. The analysis carried out in $f(T)$ gravity in Ref.~\cite{Zheng:2010am} employs the subhorizon limit $k \gg aH$ throughout its calculations in order to obtain the M\'{e}sz\'{a}ros equation. Here, we present a more general 
approach, by obtaining a generalized second order differential equation to analyse at what scales does the subhorizon limit starts to apply through the study of growth factor and growth index.

The scalar cosmological perturbation of the gravitational sector is given by the metric in the longitudinal gauge by Eq.~\eqref{eq:FLRW_metric}, obtained through the generalization of the tetrad formalism in Eq.~\eqref{eq:tetrad_choice}, where the Weitzenb\"{o}ck gauge is extended to apply at first order perturbations. Along with the perturbation of the energy momentum tensor in Eq.~\eqref{eq:pert_energymomentum}, the equations of motions~(\ref{eq:W00}--\ref{eq:Wii}) can be determined, and simplified by applying the background equations of motion. Additionally, conservation of the energy-momentum tensor provides the continuity~\eqref{eq:pert_continuity} and velocity (Euler)~\eqref{eq:pert_velocity} equations. Hence, obtaining a system of equations which can describe the evolution of scalar and matter perturbations.

The generalized M\'{e}sz\'{a}ros equation is obtained by considering CDM. Without applying the subhorizon limit at this point, all scalar modes in their spatially Fourier transformed form are expressed in terms of either higher-order time derivatives, in terms of the fractional matter perturbation $\delta_{m}$, and its derivatives using our system of equations, yielding to Eqs~(\ref{eq:varphi_expression}-\ref{eq:psi_expression}, \ref{eq:Dvarphi_expression}-\ref{eq:Dpsi_expression}, \ref{eq:DDpsi_expression}). Finally, the equation in terms of only $\delta_{m}$ and its derivatives is obtained in Eq.~\eqref{eq:meszaros_eq_in_t} where coefficients defined in Eqs~(\ref{eq:coeff_Ddeltam}-\ref{eq:coeff_deltam}) are dependent on the wave mode $k$. In  $f(\lc{R})$ gravity, using this more generalized approach in Ref.~\cite{delaCruz-Dombriz:2008ium}, it is shown that the $G_{\text{eff}}$ term is not identical to the quasi-static result, commonly reported, which was obtained by ignoring the time derivatives of  potentials.  This fact allows to greatly simplify the analysis~\cite{Tsujikawa:2007gd,DeFelice:2010aj-f(R)_theories}. 

In this work, Eqs.~(\ref{eq:coeff_Ddeltam_sub}-\ref{eq:coeff_deltam_sub}) show that the $f(T)$ subhorizon limit, presented in Ref.~\cite{Zheng:2010am}, can be obtained through this generalized approach, and, in the case of $f(T)$ models, there are no discrepancies. Moreover, we provide further analysis on what is meant by the inequality $k \gg aH$.

The five $f(T)$ teleparallel gravity models, that is Power Law, Linder, Exponential, Logarithmic and Hyperbolic-Tangent, as compiled by Ref.~\cite{Nesseris:2013jea}, provide us with prominent cases studied in the literature. For all cases, a variety of data set combinations are considered to determine constraints on $H_{0}$, $\Omega_{m0}$ and $\beta$ parameters, provided through Markov Chain Monte Carlo (MCMC) implementation in Ref.~\cite{Briffa:2021nxg}. This allows for a numerical solution of the growth factor $D(a)$ in Eq.~\eqref{eq:meszaros_in_a} and that of a constant growth index in Eq.~\eqref{eq:growth_index_differential}. By using the subhorizon limit, $k$ values  for the growth index $\gamma$, matching up to 4~d.p, have been obtained. Additionally, the $k$ value at the radiation-matter equality, $k_{\text{eq}}$, is also determined for comparison. Albeit the subhorizon limit holds below the $k_{\text{eq}}$ value in some cases, it should be noted that, for our system to hold, the minimum value should be in the matter-dominated epochs. Moreover, the consideration of higher index precision will come into play as future surveys provide an improved analysis of the growth index. As for precision up to 2 d.p., only four cases yield a result that is above the $k_{\text{eq}}$ of the respective model. These results have been summarized in Table~\ref{tab:2dp}. In particular, we provide an analysis to obtain some constraints for $\beta_{4}$ parameter in the Logarithmic model in Sec.~\ref{subsec:log}, summarized in Table~\ref{tab:log_summary}.  

\begin{table}[h!]
\footnotesize
    \centering
    \begin{tabular}{|>{\centering\arraybackslash}p{4cm}|>{\centering\arraybackslash}p{4cm}|>{\centering\arraybackslash}p{1.8cm}|>{\centering\arraybackslash}p{1.8cm}|}
    \hline
    Model  &  Data Set & $\gamma$ (2 d.p.) & $k/H_{0}$ \\ \hline\hline
    Power Law     & CC + SN + R19 & $0.54$ & $> 79.9281$\\
    Power Law &  CC + SN + R19 &$0.54$& $ > 49.1356$ \\
    Hyperbolic Tangent & CC + SN & $0.49$& $> 94.5143$ \\
    Hyperbolic Tangent & CC + SN + TRGB &$0.49$& $> 66.8440$ \\
    \hline
    \end{tabular}
    \caption{Models which attain the subhorizon growth index  $\gamma$ match up to 2~d.p. at $k > k_{\text{eq}}$. For the rest of the models, not listed here, all values above $k_{\text{eq}}$ suffices to attain the precision up to 2~d.p.}
    \label{tab:2dp}
\end{table}

\begin{table}[h!]
\footnotesize
    \centering
    \begin{tabular}{|c|c|c|}
    \hline
    Logarithmic Data Set & $\beta_{4}$ Range & $\gamma$ Range\\
    \hline \hline
    CC + SN &  $0.0476374 < \beta_{4} \ll 1146.12$ & $\gamma < 0.6908$\\
    CC + SN + R19 & $0.0516751 < \beta_{4} \ll 1056.57$ & $\gamma < 0.6860$\\
    CC + SN + HW & $0.0513575 < \beta_{4} \ll 1063.10$ & $\gamma < 0.6863$\\
    CC + SN + TRGB & $0.0482458 < \beta_{4} \ll 1131.67$ & $\gamma < 0.6900$\\
    CC + SN + BAO & $0.0351689 < \beta_{4} \ll 1552.46$ & $\gamma < 0.7089$\\
    CC + SN + BAO + R19 & $0.0329515 < \beta_{4} \ll 1656.93$ & $\gamma < 0.7127$\\
    CC + SN + BAO + HW & $0.0331934 < \beta_{4} \ll 1644.85$ & $\gamma < 0.7123$\\
    CC + SN + BAO + TRGB & $0.0339258 < \beta_{4} \ll 1609.34$ & $\gamma < 0.7110$\\
    \hline
    \end{tabular}
    \caption{$\beta_{4}$ parameter and growth index $\gamma$ constraints obtained from restrictions of Eqs~\eqref{eq:beta4_Geff}, \eqref{eq:beta4_fT} and \eqref{eq:beta4_gamma_undefined}. Minimum growth index constraint is either negative or undefined.}
    \label{tab:log_summary}
\end{table}

It would be interesting to consider a varying growth index analysis based on a constant value at high redshifts along with a value which accounts for small redshifts. Additionally, the generalized formalism presented here is a stepping stone to construct a broader framework that leads to analysis of growth in the radiation era with the considerations of anisotropies in the super-horizon limit. Finally, it is worth considering  broader classes of teleparallel theories, in particular  scalar-tensor theories such as $f(T,\phi)$, Also in these cases,  it is possible to produce second order equations of motion. Investigating  possible effects of the scalar field $\phi$ within this regime can contribute to enrich the phenomenology towards more physically reliable models.

\subsection*{Acknowledgements}

The Authors would  like to acknowledge funding from ``The Malta Council for Science and Technology'' in project IPAS-2020-007. This paper is based upon work from COST Action CA21136 {\it Addressing observational tensions in cosmology with systematics and fundamental physics} (CosmoVerse) supported by COST (European Cooperation in Science and Technology). SC acknowledges the {\it Istituto Italiano di Fisica Nucleare} (INFN) {\it iniziative specifiche} QGSKY and MOONLIGHT2.  MC  acknowledges funding by
the Tertiary Education Scholarship Scheme (TESS, Malta).

\begin{appendices}

\section{Field equations functions}\label{app:FE_functions}

The system of first order perturbation field equations~(\ref{eq:W00}-\ref{eq:Wii}) can be simplified using the continuity equation ~\eqref{eq:continuity_2}, velocity equation~\eqref{eq:velocity_2}, and gauge invariant comoving fractional matter perturbation~\eqref{eq:fractional_matter} such that each component can be expressed as a function of variables in their Fourier transform form and their higher order time derivatives:
\begin{align}
    \dudt{W}{A}{\mu}{\text{components}} = \mathcal{W}_{1}(\varphi, \dot{\varphi},\psi,\dot{\psi},\ddot{\psi}, \omega, \dot{\omega}, \delta_{m}, \dot{\delta}_{m})\,.
\end{align}
While the off-diagonal spatial-spatial component, given by Eq.~\eqref{eq:Wij}, provides an expression of $\omega$ in terms of the other variables, the rest of the components are reduced to the set of Eqs.~(\ref{eq:W00_sub}-\ref{eq:Wii_sub}) for which functions $\mathcal{F}_{i}$ ($i \in [1,4]$) are given by
\begin{align}
    \mathcal{F}_{1} &= \frac{2 \frac{k^{4}}{a^{4}} H^{2} (1+f_{T}
    -36 \frac{a^{2}}{k^{2}} H^{2} \dot{H}f_{TT}) }{\dot{H} \left(\frac{k^{2}}{a^{2}} - 3\dot{H}\right)} 
     \varphi
    + \frac{2 \frac{k^{2}}{a^{2}} (1+f_{T}) (- H^{2} + \dot{H}) }{\dot{H}}  \psi 
    + \kappa^{2} \rho \delta_{m} \nonumber \\
    & \qquad
    + \frac{6 H \left(1+f_{T} - 12 H^{2} f_{TT}\right) }{\frac{k^{2}}{a^{2}} - 3 \dot{H}} \left(\tfrac{k^{2}}{a^{2}} \dot{\psi} - \dot{H} \dot{\delta}_{m}\right)\,, \\
    \mathcal{F}_{2} &=
    \frac{6 \frac{k^{2}}{a^{2}} (1+f_{T} - 36 \frac{a^{2}}{k^{2}} H^{2} \dot{H} f_{TT})}{\frac{k^{2}}{a^{2}} - 3 \dot{H}}   ( H \varphi + \dot{\psi})
    -72 H \dot{H} f_{TT} \psi 
    -\frac{6 \dot{H} (1+f_{T} - 12 H^{2} f_{TT})}{\frac{k^2}{a^2} - 3 \dot{H}} \dot{\delta}_{m}\,, \\
    \mathcal{F}_{3} &=
    -\frac{2 \frac{k^2}{a^2} H (1+f_{T})}{ \dot{H}} \psi
    +\frac{2 \frac{k^{4}}{a^4} H (1 + f_{T} - 36 \frac{a^2}{k^2} H^2 f_{TT})}{\dot{H} (\frac{k^2}{a^2} - 3 \dot{H})} \varphi
    + \frac{6 (1 + f_{T} - 12 H^2 f_{TT} ) }{ \frac{k^{2}}{a^2} - 3 \dot{H} }
    \left(\frac{k^2}{a^2}\dot{\psi} - \dot{H} \dot{\delta}_{m} \right)\,, \\
    \mathcal{F}_{4} &= 
    \frac{k^2}{a^2} \left( (1 + f_{T})\left( 1 + \frac{H^2}{\dot{H}} - \frac{H \ddot{H}}{\dot{H}^{2}}\right) - 12 H^{2} f_{TT} 
 \right) (\psi - \varphi) \nonumber \\
 & \qquad + \left( \frac{k^2}{a^2} \frac{H}{\dot{H}} (1+f_{T}) + 3 H (1 + f_{T} - 12 H^2 f_{TT})  \right) \left(\dot{\psi} -\dot{\varphi} \right) \nonumber \\
 & \qquad - \left( 6 \dot{H} (1+f_{T}) + 9 H^2 (1 + f_{T} - 20 \dot{H} f_{TT}) - 108 H^{4}(f_{TT} - 4 \dot{H} f_{TTT})
 \right) \varphi \nonumber \\
 & \qquad  - 12 H \left(1 + f_{T} - (12 H^{2} + 9 \dot{H}) f_{TT} + 36 H^2 \dot{H} f_{TTT}\right) \dot{\psi} - 3 (1 + f_{T} - 12 H^2 f_{TT}) \ddot{\psi}\,.
\end{align}
Hence, equations for the zeroth order derivative variables can be obtained in terms of higher-order derivatives and $\delta_{m}$. One choice is that of deriving the expressions by simultaneously solving Eqs~\eqref{eq:W0i_sub} ($\mathcal{F}_{2}$) and \eqref{eq:Wi0_sub} ($\mathcal{F}_{3}$) such that
\begin{align}
    \varphi \coloneqq \mathcal{G}_{1} &= \frac{ \dot{H} (1 + f_{T}  - 36 \frac{a^2}{k^2} \dot{H}^{2} f_{TT}) (1 + f_{T} - 12 H^2 f_{TT})}
    { \frac{k^2}{a^2} H (1 + f_{T} - 12 \dot{H} f_{TT}) [ 1+f_{T} - 36 \frac{a^2}{k^2} H^2 \dot{H} f_{TT} ]} 
    \dot{\delta}_{m} \nonumber \\
    & \qquad  
    - \left( \frac{ (1 + f_{T}) }
    { H (1 + f_{T} - 12 \dot{H} f_{TT})} - \frac{  36 \frac{a^2}{k^2} \dot{H}^2 f_{TT} (1 + f_{T} - 12 H^2 f_{TT}) }
    { H (1 + f_{T} - 12 \dot{H} f_{TT})( 1 + f_{T} - 36 \frac{a^2}{k^2} H^2 \dot{H} f_{TT})} \right) \dot{\psi}\,, \\
    \psi \coloneqq \mathcal{G}_{2} &= -\frac{1 + f_{T}}{H (1 + f_{T} - 12 \dot{H} f_{TT})} \dot{\psi} + \frac{\frac{a^2}{k^2} \dot{H} (1 + f_{T} - 12 H^2 f_{TT}) }{H (1 + f_{T} - 12 \dot{H} f_{TT})} \dot{\delta}_{m}\,.
\end{align}
The rest of the field equations, along with the continuity and velocity equations and with their time derivatives, are used to express the time derivatives of variables in terms of higher-order derivatives and $\delta_{m}$ to ultimately obtain an equation as a function of $\delta_{m}$ and its derivatives:
\begin{align}
    0 = \mathcal{W}_{2}(\delta_{m}, \dot{\delta}_{m},\ddot{\delta}_{m})\,,
\end{align}
resulting in the M\'{e}sz\'{a}ros equation~\eqref{eq:meszaros_eq_in_t}. It should be noted that these equations are derived without assuming the subhorizon limit $k \gg aH$, such that we have a more general M\'{e}sz\'{a}ros equation when considering the growth of CDM.

\end{appendices}


\bibliography{sn-bibliography}

\end{document}